\documentclass{amsart} 
\usepackage{mathrsfs,amssymb,amsmath,amsfonts}
\usepackage{hyperref,graphicx,color,verbatim}

\begin{document}
\sloppypar \sloppy
\newcommand{\B}{\mbox{$\mathbb{Z}_2$}} 
\newcommand{\Z}{\mbox{$\mathbb{Z}$}}
\newcommand{\R}{\mbox{$\mathbb{R}$}}
\newcommand{\C}{\mbox{$\mathbb{C}$}}

\newcommand{\A  }{\mbox{$\textrm{Aff}(3)$}}  
\newcommand{\E  }{\mbox{$E(3)$}}  
\newcommand{\CAZ}{\mbox{$(\C \otimes \A)(\Z^3)$}}
\newcommand{\AZ }{\mbox{$\A(\Z^3)$}}
\newcommand{\CAL}{\mbox{$(\A \otimes \C \otimes \Lambda)(\R^3)$}}
\newcommand{\CL }{\mbox{$(\C \otimes \Lambda)(\R^3)$}}
\newcommand{\CZ }{\mbox{$\C(\Z^3)$}}

\newcommand{\Cl}{\mbox{$Cl(3,3,\R)$}} 

\newtheorem{theorem}{Theorem}
\newtheorem{axiom}{Axiom}
\newtheorem{thesis}{Thesis}
\newtheorem{definition}[theorem]{Definition}
\newtheorem{postulate}[axiom]{Postulate}
\renewcommand{\a}{\alpha}
\renewcommand{\b}{\beta}
\newcommand{\g}{\gamma}
\renewcommand{\d}{\delta}
\newcommand{\D}{\Delta}
\newcommand{\e}{\varepsilon}
\renewcommand{\i}{\iota}
\renewcommand{\k}{\kappa}
\renewcommand{\l}{\lambda}
\renewcommand{\L}{\Lambda}
\newcommand{\s}{\mbox{$\sigma$}}
\newcommand{\w}{\mbox{$\omega$}}

\newcommand{\pd}{\mbox{$\partial$}} 

\newcommand{\alg}[1]{\mathfrak{#1}}

\newcommand{\Ue }{\mbox{$U(1)_{\tilde{Q}}$}} 
\newcommand{\Ie }{\mbox{$\tilde{Q}$}} 

\renewcommand{\c}{\mbox{$\vec{c}$}} 

\renewcommand{\t}{\mbox{$\tau$}}  
\newcommand{\ti}{\mbox{$\tau_i$}} 
\newcommand{\hi}{\mbox{$\vec{h}_i$}}  
\newcommand{\h}{{\xi}}  
\newcommand{\hb}{{\omega}}  


\title[The SM fermions as ether excitations]{The Standard Model fermions as excitations of an ether}
\author{I. Schmelzer}
\thanks{Berlin, Germany}
\email{ilja.schmelzer@gmail.com}%
\urladdr{ilja-schmelzer.de}

\begin{abstract}
The three-dimensional complexified exteriour bundle $\mathbb{C} \otimes \Lambda(\mathbb{R}^3)$ is proposed as a geometric interpretation of electroweak doublets of Dirac fermions. The Dirac equation on this bundle allows a staggered discretization on a three-dimensional scalar complex lattice field $\mathbb{C}(\mathbb{Z}^3)$. 

The three-dimensional affine group $\mathrm{Aff}(3)$ is proposed as a geometric interpretation for the matrix defined by the three generations, with the three quark doublets and a lepton doublet in each generation. The corresponding lattice space $(\mathrm{Aff}(3) \otimes \mathbb{C})(\mathbb{Z}^3)$ allows a simple condensed matter (ether) interpretation. A possibility to construct anticommuting fermion operators based on canonical quantization is given.

A set of axioms of the gauge action is defined such that the gauge action of the SM gauge group $SU(3)_c\times SU(2)_L\times U(1)_{Y}$ is the maximal possible one. This includes preservation of Euclidean symmetry and the symplectic structure, anomaly freedom, ground state neutrality, and the possibility of realization on the lattice. Strong interactions are realized as Wilson-like gauge fields, weak gauge fields as effective gauge fields describing lattice deformations, and the EM field as a combination of above types. 

Some possibilities to explain the qualitative characteristics of the mass parameters are discussed. 

\end{abstract}

\maketitle


\section{Introduction}

The standard model of particle physics is a remarkable success of modern physics. Developed in the seventies of the last century, it has been able to predicts thousands of empirical data obtained up to now. But it cries for explanation. Nor the number and structure of the fermions, nor the gauge group $SU(3)_c\times SU(2)_L\times U(1)_{Y}$, nor its action on the fermionic sector is in any way explained. One can live with the idea that the mass parameters are arbitrary, similar to material properties of various materials. But there are a lot of remarkable qualitative exact properties: Three generations with exactly the same gauge action, the exact inertness of right-handed neutrinos, independence of the $SU(3)_c$ action on weak isospin and chirality, independence of the $SU(2)_L$ action on color, lepton and baryon charge, as well as the independence of the $U(1)_{\gamma}$ action on color and chirality. And at least some qualitative properties of the masses also cry for a qualitative explanation: The masslessness of the strong and the EM field and the extremely small masses of the neutrinos in comparison with the other fermions.

One attempt to find such explanation is, of course, string theory. But if one looks at the current state, at least as described by Woit \cite{Woit}, it seems naive to hope for an explanation of all these properties from string theory. So alternatives to string theory deserve some consideration.

Of course, one would not expect that such an alternative theory would follow all the actually fashionable ideas -- or prejudices -- about our world. One would expect that some ideas, even ideas which seem unquestionable today, have to be given up. In modern physics, not only the dimension of space is questioned by string theory, but even fundamental philosophical concepts like realism are rejected. So, it seems, there should be no objection against even extremely strange theories, and there should be no theoretical principle which is not open to rejection by such an alternative theory, except the basic principle of agreement with observation. 

But I'm afraid that the basic ideas of the approach presented here are an exception, that they are too offensive for many modern physicists. The point is that these ideas are not new at all, but quite old, pre-relativistic, ideas, ideas rejected by the relativistic revolution, and, therefore, discredited. The idea that the spacetime is, in fact, twenty-six or eleven-dimensional, with most of these dimensions remaining hidden is today, mainstream, and probably even the idea that the dimension of spacetime is non-integer would meet less prejudice than the idea that spacetime splits, on a hidden, fundamental level, into a three-dimensional Euclidean space and one-dimensional time. And interpreting fermion fields in terms of tetrads or Grassmann fields is clearly more popular than an interpretation in terms of waves of some condensed matter, which could be named, in the pre-relativistic tradition, the ether. 

The motivations for this strong prejudice are at least partially outside the domain of physics. 

There is the sociological problem that these ideas have been discredited by various ``ether theorists'' -- cranks who claim to have found logical contradictions in special relativity and similar nonsense -- and the anti-semitic tradition of Nazistic ``German science''. A theory with such friends does not need enemies. 

A psychological point may be not unimportant too: The relativistic arguments against these ideas are one of the first things one learns about modern physics. And this early lecture is a quite impressive one, it makes our world much more strange and fascinating. This fascination may have been part of the motivation to become a physicists. At least the author's interest in physics has been influenced by this strangeness. 

The modern organization of science has to be mentioned too -- ``publih or perish'' and survival on short-time grants lead to concentration in a few mainstream directions like string theory: Following the mainstream gives more journals to publish, more conferences, more conference proceedings, more readers of your papers who possibly cite them, more grants and positions, more of everything important for a scientific career -- except truth which may be hidden somewhere else. 

In comparison with these extra-physical problems, the physical objections do not seem that impressive. Of course, there has to be a mechanism which explains why absolute time and absolute distances are unobservable, what distorts the clocks and rulers. But this is an easy exercise in comparison with hiding many spacetime dimensions. Given the violation of Bell's inequality, the motivation of introducing a hidden preferred frame is, in fact, of almost ultimate strength -- all one has to presuppose is realism.

But the very task of explanation of the particle content of the SM adds independent motivation: There are isospin rotations, which can be combined with spinor rotations into a representation of $SO(3)$, so that a three-dimensional geometric interpretation of electroweak pairs becomes possible. There are also three generations and three quark colors. For me, a standard model with eleven generations and eleven colors would be a motivation for taking string theory seriously. But our world is different, and the idea that our four-dimensional spacetime splits into some three-dimensional part we can name ``space'' and a one-dimensional part we can name ``time'', such that the three-dimensional part becomes geometrically associated with the number of generations and colors, seems more plausible to me. 

The aim of this text is to provide an ether model which realizes these possibilities. Each fermion generation, each quark color, and each of the generators $I_i$ of isospin rotations becomes associated with a direction in space. 

The focus of interest of this text is on fermions. So many important questions are not considered here. In particular the explanation of relativistic symmetry is done in a corresponding theory of gravity, which derives the EEP and gives the Einstein equations of GR is a natural limit in \cite{GLET}. The completely independent argumentation for a hidden preferred frame following from the violation of Bell's inequality \cite{Bell} is not considered here. 

On the other hand, the model provides natural explanations for the properties of the gauge action on the SM fermions, so it is considered here too. 

\subsection{Basic ideas}

As the three-dimensional geometric interpretation of the SM fermions, the bundle \CAL\/ is proposed. The bundle \CL\/ is identified with a doublet of Dirac fermions.\footnote{Independent of this paper, three-dimensional geometric fermions have been proposed by Daviau \cite{Daviau}, and the idea that geometric fermions may be used to describe electroweak doublets has been proposed by Hestenes \cite{Hestenes}.}
For this purpose, the three-dimensional geometric Dirac operator on \CL,  an analogon of the Dirac-K\"{a}hler operator \cite{Kaehler} on $(\C \otimes \Lambda)(\R^4)$, will be used. This three-dimensional Dirac operator is sufficient to define the Dirac matrices $\a^i$. There are also natural operators $I_i$ as well as $\b=\g^0$ on \CL. The Dirac equation can be defined in its original Dirac form $i\partial_t\psi = H \psi$, as an evolution equation on \CL. This equation contains eight complex fields and correspondingly describes a doublet of Dirac particles. 

There are $3\cdot(3+1)$ such components in \CAL, corresponding to the matrix representation $(a^i_\mu)\in\A$ of an affine transformations. Each of these components will be identified with an electroweak doublet: The upper index $i$ denotes the generation, $\mu=0$ the leptonic sector, $\mu>0$ the quark sector, where the three positive values $\mu\in\{1,2,3\}$ define the three quark colors.

In analogy with the staggered discretization of the four-dimensional bundle $(\C \otimes \Lambda)(\R^4)$, there exists also a staggered discretization of the three-dimensional Dirac operator. It lives on a three-dimensional spatial lattice $\Z^3$, with time left continuous. It is a staggered discretization, with only one complex component on each lattice node, and eight different types of lattice nodes. Similar to the four-dimensional staggered discretization of $\Lambda(\R^4)$ on $\Z^4$ (see \cite{Banks,Banks1,Becher,BecherJoos,Rabin,Susskind}), it is a doubler-free discretization of the Dirac equation on \CL. In other words, we obtain a lattice evolution equation on a three-dimensional lattice \CZ, which gives, in the continuous limit, two Dirac fermions.

\begin{figure}
\includegraphics[angle=0,width=0.8\textwidth]{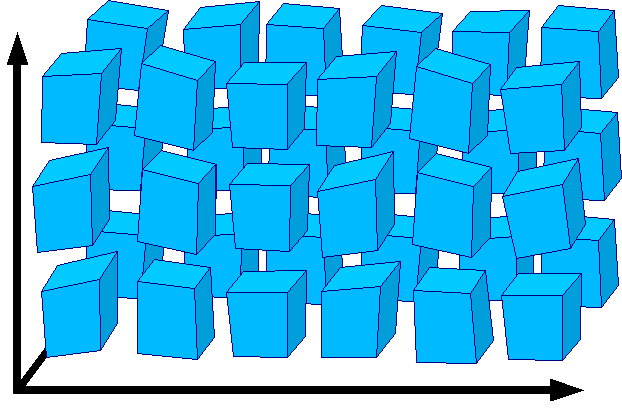}
\label{fig:lattice}\caption{The space \CAZ\/ of the lattice model suggests an interpretation as the phase space (with configuration space \AZ) of a three-dimensional lattice of deformable three-dimensional cells. The configuration of each cell is described by an affine transformation from a standard reference cell.}
\end{figure}

For all SM fermions (the bundle \CAL) we obtain a first order lattice equation on \CAZ. This lattice space allows a physical interpretation as the phase space for a three-dimensional lattice of elementary cells, where the state of each cell is described by a single affine transformation from a standard reference cell (see figure \ref{fig:lattice}).

This physical interpretation gives us two important structures: First, a symplectic structure of the phase space, second, a natural action of the Euclidean group \E. These structures may be used to restrict the action of the gauge group. For a compact gauge group, we can always construct a preserved Euclidean metric on the phase space, which, together with a preserved symplectic structure, allows to construct a preserved complex structure. Thus, preservation of the symplectic structure requires unitarity of the gauge groups.

The left action of \E\/ on \AZ\/ transforms the lattice as a whole. The requirement of preserving this symmetry for the gauge groups consists of two parts: 

\begin{itemize}
\item To commute with rotations, gauge groups have to preserve generations and to act on all three generations in the same way. This holds for all SM gauge fields.
\item To commute with translations, one direction (one component in each generation) has to be preserved. All SM gauge fields leave right-handed neutrinos and their antiparticles invariant, thus, a common invariant direction exists in the SM.
\end{itemize}

Thus, after an appropriate identification of the invariant direction, all SM gauge fields preserve \E\/ symmetry. 

The lattice theory also leads to another important restriction for the gauge fields: We have to define an appropriate lattice model for the gauge fields. A well-known way to put gauge fields on the lattice are Wilson gauge fields. Their modification to a three-dimensional lattice with continuous time is straightforward. But Wilson gauge fields cannot act in a nontrivial way inside the doublets \CL, because these are represented on the lattice as \CZ, which leaves $U(1)$ as the maximal possible Wilson gauge field. Thus, Wilson gauge fields have the same charge on all parts of a doublet. The maximal group of Wilson gauge fields compatible with this restriction, \E\/ symmetry, and symplectic structure is $U(3)\cong SU(3)_{QCD}\times U(1)_B$.

There exists another modification of the lattice equations which, in the large distance limit, leads to a gauge-like interaction term for fermions.  It describes correction terms for lattice deformations.\footnote{The idea that gauge fields may be obtained as effective fields in condensed matter theory to describe various types of lattice defects is quite old, see, for example, \cite{Kadic}.} The coefficients  modify the Dirac equation on \CL\/ and depend only on the geometry of the lattice. So the resulting equation preserves doublets \CL, and acts identically on all doublets. One can construct generators as operators commuting with the lattice Dirac equation, which correspond to the basic lattice shifts on the staggered lattice. They appear to be generated by $I_i$ and $\g^5$. This gives the maximal gauge group $U(2)_L\times U(2)_R$, which has to be reduced to allow for translational symmetry. One choice of the maximal group of this type, compatible with \E\/ symmetry, is generated by chiral $U(2)_L\cong SU(2)_L\times U(1)_L$, and a vector field \Ue\/ with charge $\Ie = I_3-\frac{1}{2}$.

The EM field does not fit into any of the two types. But it can be constructed as a combination of them, by the simple formula $Q=2I_B+\Ie$. Thus, our two types of gauge-like lattice fields are already sufficient to construct all SM gauge fields. 

Based on arguments from condensed matter theory, one can justify the further requirement that the gauge group should be special, that means, that the charges have zero trace. This excludes the field $U(1)_B$, which would, otherwise, be allowed. What remains as the maximal possible gauge group $G_{max}$ is the SM gauge group $G_{SM}$, together with only one additional gauge field --- an ``upper axial gauge field'', with charge $I_{U} = \g^5(I_3-\frac{1}{2})$, which acts on upper quarks and leptons. This extension of the SM is anomalous, thus, if we add, as a last conditions, anomaly freedom, we obtain $G_{SM}$ as a maximal anomaly-free subgroup of $G_{max}$. Thus, one can, essentially, compute the SM gauge group and its action on the fermions from first principles.

Of course, there are a lot of things left to future research. We have not considered yet the Higgs sector, and there are only a few vague ideas about mass terms. They break the left \E\/ symmetry preserved by the gauge action. Thus, to describe them, we need some spontaneous \E-symmetry breaking. Once the broken symmetry is \E, it is not clear if we need a separate Higgs sector at all. This has to be left to future research.

What about quantization? The first problem is fermion quantization. The model is based on classical, commuting c-number fields, which define the lattice \CAZ. There are no Grassmann variables as in the Berezin approach to fermion quantization. Thus, we need a completely different quantization scheme for fermions.

Such an alternative is proposed in section \ref{fermionQuantization}. It is based on canonical quantization of a field with \B-symmetric degenerated vacuum state. The lowest energy states of such a field define a \B-valued (spin) field, but yet with commuting operators on different lattice nodes. Fortunately, the algebra of (anticommuting) lattice fermion operators and the algebra of (commuting) spin field operators appear to be isomorph. Unfortunately, the isomorphism is nonlocal and depends on an order between the lattice nodes. But one can motivate a particular choice of this order. For this choice, one obtaines the staggered lattice Dirac operator (exactly in dimension one, approximately in higher dimensions) from a much simpler, non-staggered, direction-independent spin-field Hamilton operator, which has been obtained from canonical quantization.

The resulting lattice gauge fields appear as effective fields. Their quantum effects are similar to phonons in condensed matter theory. Therefore, they are ``quantized'' already by the quantization of the fundamental (high energy) degrees of freedom --- the states of the elementary cells (and the material between them). Instead of a separate quantization procedure for these fields, the effective quantum properties of these fields have to be derived from the fundamental quantum theory. The details have to be worked out by future research. Nonetheless, some differences are obvious. Especially, different gauge-equivalent configurations describe different physical states. Therefore, the standard rejection of anomalous gauge fields based on the BRST quantization procedure, which requires strong gauge invariance, seems premature. That means, the additional field $U(1)_{U}$ may as well appear in the final theory.

The gravitational field describes, as an effective large distance field, density, average velocity, and the stress tensor of the medium. So it does not appear on the microscopic level, and, as well, does not require a separate quantization procedure. This is not considered here. The classical theory is described in \cite{GLET}. The consideration of quantum effects, which have to be handled similar to quantum effects of usual condensed matter, is left to future research.

\section{Geometric interpretation of SM fermions}

Let's consider at first the geometric interpretation of the SM fermions as sections of the bundle \CAL. We do not consider mass terms. Without the mass terms, the three generations of SM fermions appear completely identical, simply as three identical copies of the same representation of the SM gauge group $SU(3)_c\times SU(2)_L\times U(1)_Y$.

The group \A\/ is the group of three-dimensional affine transformations $y^i=a^i_j x^j + a^i_0$ on $\R^3$. Each $a^i_\mu$ can be identified with an electroweak doublet of the SM according to the following simple rules: The upper index $i$, $1\le i \le 3$, defines the generation. The translational components $a^i_0$ is identified with the leptonic sector. The linear part $a^i_j$, $j>0$, is identified with the quark sector. In this case, the lower index $j$, $1\le j\le 3$, denotes the color of the quark doublet. 

This identification of the $3\times(3+1)$ SM doublets with a $3\times(3+1)$ affine matrix may be considered, up to now, as pure numerology. But it defines a natural action of the Euclidean group \E, by multiplication from the left. This action commutes, as we will see, with all SM gauge fields and plays an important part in our computation of the SM gauge action.

Each electroweak doublet is defined by the bundle \CL. It is assumed here that right-handed neutrinos exists, so that neutrinos form usual Dirac particles. Thus, qualitatively there is no difference between electroweak quark doublets and electroweak lepton doublets: Above contain two Dirac particles. The bundle \CL\/ consists of three-dimensional complex inhomogeneous differential forms
\begin{equation}
\begin{split} \Psi &= \sum_{\k_i\in\{0,1\}} \psi_{\k_1\k_2\k_3}(x) e^{\k_1\k_2\k_3}\\
 &= \psi_{000}(x)\\
 &\quad + \psi_{100}(x)\, dx^1 + \psi_{010}(x)\,dx^2+\psi_{001}(x)\,dx^3 \\
 &\quad + \psi_{110}(x)\,dx^1 \wedge dx^2+\psi_{011}(x)\,dx^2\wedge dx^3+\psi_{101}(x)\,dx^1\wedge dx^3 \\
 &\quad + \psi_{111}(x)\, dx^1 \wedge dx^2 \wedge dx^3.
\end{split}
\end{equation}

Thus, we have $1+3+3+1=8$ complex functions, which gives two Dirac fermions. This allows a physical interpretation in terms of a standard model electroweak doublet. The use of a three-dimensional bundle is essential: In Minkowski spacetime, we have only the bundle $\C\times\Lambda(\R^4)$, with the Dirac-K\"{a}hler equation \cite{Kaehler}. But this equation describes four Dirac fermions.

On the external bundle $\Lambda(\R^d)$ exists a natural geometric Dirac operator $D$ as a square root of the Laplace operator $\Delta = D^2$. For a general metric, the definition is given in appendix \ref{DiracCurved}. In the Euclidean case $g_{\mu\nu}=\delta_{\mu\nu}$, this Dirac operator has the form
\begin{equation}
 D = d + d^* = -i\a^i\pd_i.
\end{equation} 
with operators $\a^i$, which fulfill the anticommutation relations $\{\a^i,\a^j\} = 2\delta^{ij}$.  Now, together with the skew-symmetric $\a^i$, it is useful to consider also corresponding symmetric operators $\b^i$. They may be defined analogically as
\begin{equation}
 d - d^* = -i\b^i\pd_i.
\end{equation} 
Together, they define a set of generators of $M_{2^d}(\R)\cong\textit{Cl\;}^{d,d}(\R)$:
\begin{equation}
  \{\a^i,\a^j\} = 2\d^{ij}, \;
  \{\a^i,\b^j\} = 0, \;
  \{\b^i,\b^j\} = -2\d^{ij}.
  \label{eq:aibi}
\end{equation}
For $d=3$, the explicit representation of the matrices $\a^i,\b^i$ is defined by:
\begin{align}\label{def:representation}
-i\a^i\pd_i\Psi &= \begin{pmatrix} 
      0 & -\pd_3 & -\pd_2 &      0 & -\pd_1 &      0 &      0 &      0 \\
 +\pd_3 &      0 &      0 & -\pd_2 &      0 & -\pd_1 &      0 &      0 \\
 +\pd_2 &      0 &      0 & +\pd_3 &      0 &      0 & -\pd_1 &      0 \\
      0 & +\pd_2 & -\pd_3 &      0 &      0 &      0 &      0 & -\pd_1 \\
 +\pd_1 &      0 &      0 &      0 &      0 & +\pd_3 & +\pd_2 &      0 \\
      0 & +\pd_1 &      0 &      0 & -\pd_3 &      0 &      0 & +\pd_2 \\
      0 &      0 & +\pd_1 &      0 & -\pd_2 &      0 &      0 & -\pd_3 \\
      0 &      0 &      0 & +\pd_1 &      0 & -\pd_2 & +\pd_3 &      0
\end{pmatrix} 	
\begin{pmatrix} 
\psi_{000}\\
\psi_{001}\\
\psi_{010}\\
\psi_{011}\\
\psi_{100}\\
\psi_{101}\\
\psi_{110}\\
\psi_{111}
\end{pmatrix}\\ 
-i\b^i\pd_i\Psi &= \begin{pmatrix} 
      0 & +\pd_3 & +\pd_2 &      0 & +\pd_1 &      0 &      0 &      0 \\
 +\pd_3 &      0 &      0 & +\pd_2 &      0 & +\pd_1 &      0 &      0 \\
 +\pd_2 &      0 &      0 & -\pd_3 &      0 &      0 & +\pd_1 &      0 \\
      0 & +\pd_2 & -\pd_3 &      0 &      0 &      0 &      0 & +\pd_1 \\
 +\pd_1 &      0 &      0 &      0 &      0 & -\pd_3 & -\pd_2 &      0 \\
      0 & +\pd_1 &      0 &      0 & -\pd_3 &      0 &      0 & -\pd_2 \\
      0 &      0 & +\pd_1 &      0 & -\pd_2 &      0 &      0 & +\pd_3 \\
      0 &      0 &      0 & +\pd_1 &      0 & -\pd_2 & +\pd_3 &      0
\end{pmatrix}	
\begin{pmatrix} 
\psi_{000}\\
\psi_{001}\\
\psi_{010}\\
\psi_{011}\\
\psi_{100}\\
\psi_{101}\\
\psi_{110}\\
\psi_{111}
\end{pmatrix} 
\end{align}
The last Dirac operator $\b=\g^0$ can be obtained now as
\begin{equation}
	\b = \g^0 = \b^1\b^2\b^3\a^1\a^2\a^3 = \a^1\b^1\a^2\b^2\a^3\b^3,
	\label{eq:betafromaibi}
\end{equation}
and appears to be a diagonal operator, which measures the \B-graduation of $\Lambda(\R^3)$. The matrices $\a^i,\b$ define a representation of the Dirac algebra
\begin{equation}
	\{\a^i,\a^j\}=2\d^{ij};\qquad\{\a^i,\b\}=0;\qquad (\a^i)^2=\b^2=1.
	\label{eq:DiracAlgebra}
\end{equation}
For the (massless) Dirac equation we prefer to use the original form, as proposed by Dirac, with the operators $\a^i$:
\begin{equation}
 i\pd_t \Psi = -i\a^i\pd_i\Psi.
\end{equation} 
The operators $I_i$ defined by
\begin{equation}
 2i\varepsilon^{ijk}I_i = \b^j\b^k,
\end{equation} 
define a representation of the isospin algebra $\mathfrak{su}(2)$. We identify them with the (weak) vector isospin $I_i = \tau^i_L + \tau^i_R$. The $I_i$ commute, as they should, with the Dirac equation as well as with $\g^0$. Thus, the operator $I_3$ may be used to split the bundle \CL\/ into two parts with eigenvalues $I_3=\pm\frac{1}{2}$, so that each of the parts contains a full representation of the Dirac algebra.

An interesting question is how the spinor representation $\sigma^{ij}=\a^i\a^j$ on the Dirac particles is connected with the representation $\mathfrak{so}(3)$ of geometric rotations of the bundle \CL. The answer is that geometric rotations are generated by the operators $\omega^{ij}$ defined by
\begin{equation}
 \omega^{ij} = \a^i\a^j - \b^i\b^j = \sigma^{ij} - 2i\varepsilon^{ijk}I_i.
\end{equation} 
Thus, the true, geometric rotations of our geometric interpretation are a combination of spinor rotations and isospin rotations. The operator $\g^5=-i\a^3\a^2\a^1$ turns out to be the (modified) geometric Hodge $\ast$ operator (\ref{astdef}).

\subsection{Symplectic structure}\label{complex}

We have a complex structure in our geometric interpretation. Now, every complex structure defines a natural symplectic structure $\omega=dz\wedge d\bar{z}$. We know that all the SM gauge groups are unitary groups, thus, they preserve the complex structure. As a consequence, they also preserve the symplectic structure. Therefore, we can postulate the following:

\begin{postulate}\label{postulate:symplectic}
All gauge fields preserve the symplectic structure derived from the complex structure of \CL.
\end{postulate}

The question we want to consider here is if we really need the complex structure. May be the symplectic structure is already sufficient?  Or do we obtain, in this way, some additional gauge fields? No, at least as long as we consider only compact gauge groups. For compact gauge groups, we have the invariant Haar measure $d\mu(g)$, and it has a finite norm. This allows to construct, for a given action of a compact group, an invariant Euclidean norm $\langle.,.\rangle$. All we have to do is to start with an arbitrary norm $\langle.,.\rangle_0$ and to compute the average of the Haar measure:
\begin{equation}
 \langle a,b \rangle = \int \langle ga,gb \rangle_0 d\mu(g).
\end{equation} 
The resulting Euclidean distance $\langle.,.\rangle$ is already preserved by the gauge group action. Once we have a preserved Euclidean metric  $\langle.,.\rangle$  together with a preserved symplectic structure $\omega$, we can already construct a preserved complex structure by the rule
\begin{equation}
 \omega(a,ib) = \langle a,b \rangle.
\end{equation} 
As a consequence, our postulate \ref{postulate:symplectic} is sufficient to restrict the gauge group to an unitary group. Thus, in the geometric interpretation we can forget about the complex structure and restrict ourself to the symplectic structure, that means, we can interpret the space \CAL\/ as a phase space.

\subsection{Euclidean symmetry}

On \A, we have a well-defined left action of the Euclidean symmetry group $\E\subset\A$. 

The action of the rotation group $O(3)\subset\E$ extends immediately to \CAL\/ as
\begin{equation}
\omega: \Psi^i_\mu \to \omega^i_j\Psi^j_\mu.
\end{equation} 
In terms of our interpretation, these rotations rotate the three generations of the SM. Now, all SM gauge groups preserve generations. (Remember that we consider here the massless case, thus, define generations not in terms of mass eigenstates, but in such a way that they contain electroweak doublets completely.) Moreover, they act on the different generations in exactly the same way. As a consequence, they commute with the action of our group of rotations $O(3)$.

Let's extend now the action of the subgroup of translation $T^3\subset\E$ on \CAL. For this purpose, we have to define a shift operator
\begin{equation}
 t: \Psi^i_0 \to \tau(t^i)\Psi^i_0,
\end{equation}
where $\tau(t): \Psi\to\Psi'$ defines a scalar shift operator on \CL. This shift is an additive action of $\R$ on $\CL$, and it should not depend on $x$. Therefore, it is uniquely defined by a single shift vector $\c=(c_\k)\in\C^8$ as
\begin{equation}
 \c = \tau(1)\Psi-\Psi,
\end{equation} 
which we name the ``direction of translation''. After this, translations are defined as $\tau(t)\psi_\k\to\psi_\k+tc_\k$ for all $\k$, and we have extended the definition of translations from \A\/ to \CAL. 

In our interpretation, translations act, by shifts, only on the leptonic doublets. Once we already have found that rotations commute with all gauge groups, it would be nice to have a similar property for translations too. So, what does it mean for the gauge groups to commute with translations? The answer is simple -- the gauge groups have to leave the translational direction \c, which is located in the leptonic sector, invariant:
\begin{equation}\label{t-invariance}
 [g,\tau(t)] = 0 \qquad \Leftrightarrow \qquad g \c = \c
\end{equation}
Now, the leptonic sector of the SM contains directions which are left invariant by all SM gauge fields -- the right-handed neutrinos and their antiparticles. Thus, if we identify the direction of translation \c\/ in such a way, that it is inside the right-handed neutrino sector, then all SM gauge fields preserve translational symmetry too.

Thus, with an appropriate definition of the direction of translation \c, all SM gauge fields preserve the complete \E\/ symmetry. This property of the SM gauge fields we use in the following as a postulate:

\begin{postulate} \label{postulate:Euclidean} All gauge fields preserve the \E\/ symmetry defined by the left action of \E\/ on \A.
\end{postulate}

Note that this observation gives our \E\/ symmetry large explanatory power. It explains why all SM gauge fields preserve generations and act in the same way on the three generations. Moreover, it excludes a lot of very interesting natural and symmetric extensions of the SM:

\begin{itemize}

\item The extension of $SU(3)_c$ to $SU(4)_c$ with lepton charge as a forth color, which is part of the Pati-Salam extension of the SM \cite{PatiSalam},

\item the left-right-symmetric extension of $SU(2)_L\times U(1)_Y$ to $U(1)_{B-L}\times SU(2)_L\times SU(2)_R$, which is also part of the Pati-Salam extension of the SM \cite{PatiSalam},

\item and all GUTs which use at least one of these extensions as a subgroup, especially $SO(10)$ GUT. 

\end{itemize}

Indeed, all these extensions of the SM act on right-handed neutrinos in a nontrivial way, and, moreover, they also leave no other direction invariant. As a consequence, they cannot commute with any choice of the direction of translation \c, thus, cannot commute with Euclidean translations.

Nonetheless, these principles are not yet sufficient to compute the SM gauge group. There remain interesting nontrivial extensions like $SU(5)$ GUT \cite{su5} or chiral color with $SU(3)_L\times SU(3)_R$ instead of $SU(3)_c$ \cite{chiralColor}. 

\section{The lattice Dirac operator}\label{doubling}

Let's consider now a discretization of our Dirac equation in space, leaving time continuous. Using naive central differences, we obtain the following lattice equation:
\begin{equation}
	i\pd_t \psi_\k(n)   \ =\ 
		\sum_i -i(\alpha^i)_\k^{\k'} (\psi_{\k'}({n+h_i})-\psi_{\k'}({n-h_i})) 
		\label{eq:DiracLattice}
\end{equation}
on the lattice space $\Omega=\C^8(\Z^3)$. 

It is easy to see that this lattice equation contains eight doublers. Indeed, let's consider eight so-called ``staggered'' sublattices, labelled by $\l=(\l_1,\l_2,\l_3)\in\{0,1\}^3$, defined by the condition 
\begin{equation}
 \Omega^\l = \{\psi_\k(n)| n = \k+\l \;\;\textrm{mod}\;\; 2\}
\end{equation}
so that  $\Omega=\sum_\l \Omega^\l$. It is easy to see that the naive lattice Dirac equation (in our representation \eqref{def:representation}) preserves the decomposition into the staggered sublattices. As a consequence, it is easy to get rid of the doublers, and sufficient to preserve only one of the eight sublattices $\Omega^{000}$, with $\l=\{0,0,0\}$. Thus, our staggered sublattice is defined by the condition
\begin{equation}\label{eq:DiracStaggered}
 n = \k \;\;\textrm{mod}\;\; 2.
\end{equation} 
This doubler-free lattice equation (\ref{eq:DiracLattice}),(\ref{eq:DiracStaggered}) can be obtained from a much more genereal, geometric construction, which is presented in appendix \ref{DiracLatticeCurved}. It is the same geometric construction, which gives, in the case of the four-dimensional Dirac-K\"{a}hler equation \cite{Kaehler} on the spacetime bundle $\Lambda(\R^4)$, the Kogut-Susskind staggered fermions \cite{Banks1,Kogut,Susskind} in lattice gauge theory (see \cite{Banks,Becher,BecherJoos,GuptaLattice,Rabin}).

Now, it is interesting to see what happens with the other operators we have defined in the continuous limit. On $\Omega$, the operators $\a^i,\b^i,I_i,\g^5$ and the shift operators $\tau_i: \Psi(n)\to\Psi(n+h_i)$ are well-defined. Unfortunately, they do not preserve the decomposition into staggered subspaces. Fortunately, there are natural replacements for these operators, which already preserve $\Omega^{000}$. Instead of the generators $\a^i,\b^i$ of $\textit{Cl}^{3,3}(\R)$ we can use the following replacements:
\begin{equation}
 \tilde{\a}^i = \a^i\tau_i, \qquad  \tilde{\b}^i = \b^i\tau_i.
\end{equation} 
For the other operators $I_i,\g^5$ we can use the same formulas we have used in the continuous limit to compute them: 
\begin{equation}
 \tilde{\g}^5=-i\tilde{\a}^3\tilde{\a}^2\tilde{\a}^1 = \g^5\tau_1\tau_2\tau_3
\end{equation} 
\begin{equation}
 2i\varepsilon^{ijk}\tilde{I}_i = \tilde{\b}^j\tilde{\b}^k = \b^j\b^k\tau^j\tau^k
\end{equation} 
Now, the operators $\tilde{\g}^5$ and $\tilde{I}_i$ generate an interesting group $\mathcal{A}$ of operators associated with lattice shifts:
\begin{theorem}\label{th:shiftAlgebra} The group $\mathcal{A}$ of operators generated by $\tilde{\g}^5$ and $2\tilde{I}_i$ has the following properties:
\begin{itemize}
 \item It preserves the staggered subspaces $\Omega^\l$.
 \item It preserves the massless lattice Dirac equation.
\footnote{Note one advantage of using the original form $D=\a^i\pd_i$ of the Dirac equation: $\g^5$ does not anticommute, but commute with the (massless) Dirac equation.}
 \item There exists an epimorphism $\pi: \mathcal{A} \to \Z^3$ named ``underlying shift operator''.
 \item $\mathrm{Ker} \pi \cong \{\pm 1\}$ and acts by pointwise multiplication.
 \item $\forall a\in A: a^2 = (\pi a)^2$. 
\end{itemize}
\end{theorem}
For a shift operator $\tau\in\Z^3$, the equation $\pi(\tilde{\tau})=\tau$ defines the operator $\tilde{\tau}$ modulo its sign.

In the continuous limit, the subgroup of $\mathcal{A}$ generated by even shifts becomes irrelevant. The corresponding  factorgroup consists of the following operators: $\tilde{\mathcal{A}} = \{\pm1, \pm2I_i, \pm\g^5, \pm2I_i\g^5\}$. This set of operators generates the Lie algebra of the group $U(2)_L\times U(2)_R$, which plays an important role in the following. Especially it contains the weak gauge group $SU(2)_L$.

\subsection{The lattice model}

Let's forget, for some time, about the staggered character of the lattice Dirac equation. Then, the lattice space of the discretization $\Omega^{000}$ is simply \CZ, with a single complex number on each lattice node. For all SM fermions, we obtain the lattice space \CAZ. 

Note also that we have a first order lattice equation on it. This suggests an interpretation of \CAZ\/ as a phase space of some physical system:
\begin{equation}
\psi^i_\mu(n) = \varphi^i_\mu(n) + i \pi^i_\mu(n).
\end{equation}
with configuration variables $\varphi^i_\mu(n): \Z^3\to\A$ and momentum variables $\pi^i_\mu(n)$. On the phase space \CAZ\/ we have the standard symplectic structure
\begin{equation}\label{symplectic}
\omega = \sum_{i,\mu,n} d \varphi^i_\mu(n) \wedge d \pi^i_\mu(n).
\end{equation}

Now, the configuration space \AZ\/ allows a natural interpretation as a regular lattice of deformable cells in $\mathbb{R}^3$ (see figure \ref{fig:lattice}): The state of each cell is described by an affine transformation from a standard reference cell. This reference cell is assumed to be located in the origin.

Now, to have such a simple model is, of course, nice and beautiful. But is it only an otherwise useless toy, or is it helpful to explain the physics of the SM? We want to show here that this model has physical importance.

First, of course, this model gives the symplectic structure, which we have used in section \ref{complex} to derive unitarity. Thus, the model allows to explain our postulate \ref{postulate:symplectic}.

But it seems helpful to explain Euclidean symmetry too. Of course, Euclidean symmetry is not a property of the full SM, where the mass matrices break this symmetry. Thus, we need some spontaneous symmetry breaking to explain the SM masses. Nonetheless, the lattice model allows to answer the following simple question: Why do we have to use the left action of \E\/ on \A, instead of the right or adjoint action? Abstract group theory remains silent about this. Instead, for a lattice of deformed cells, we can look what happens with the lattice if we apply the different actions of \E:

Let's consider an almost regular lattice. Then we have approximately
\begin{equation}
	\varphi^i_j(n) \approx \delta^i_j, \qquad \varphi^i_0(n) \approx n_ih,
	\label{eq:locationInSpace}
\end{equation}
Now, we see that the left action of a rotation rotates the lattice as a whole, including the shifts $\varphi^i_0(n)$. Instead, the right action of a rotation leaves the cells on their places $n_ih$ and rotates them around these places. This, obviously, modifies the geometric relations between neighbour cells.  Only the left action rotates the lattice as a whole, leaving the local geometry unchanged. Thus, the left action is much more likely to be a symmetry of the theory --- something which cannot be seen otherwise.  In this sense, our cell model is useful to explain our postulate \ref{postulate:Euclidean} as well.

But the most important consequence of the lattice model is that we can apply now condensed matter theory. Especially we can, in the large distance limit, define density, velocity, and a stress tensor, and postulate continuity and Euler equations. But this is what we need to incorporate gravity into the model. A metric theory of gravity with GR limit, based on such an ``ether concept'', has been proposed in \cite{GLET}.

\section{Lattice gauge fields}

While our postulates \ref{postulate:symplectic} and \ref{postulate:Euclidean} impose strong restrictions for the gauge group of the SM, we are yet far away from computing the SM gauge group. There are, yet, gauge groups much larger than $SU(3)_c\times SU(2)_L\times U(1)_Y$, which are compatible with our postulates.

But the consideration of the lattice theory allows to impose another type of restrictions: It should be possible to ``put the gauge action on the lattice''. We will see that this gives the additional restrictions, which allow to compute the SM gauge group almost exactly. 

\subsection{Strong fields as Wilson gauge fields}

The classical way to incorporate gauge fields into a lattice theory are Wilson gauge fields. The classical formalism of Wilson gauge fields, even if it was developed for spacetime lattices $\Z^4$ instead of our lattice of cells $\Z^3$, needs only a sufficiently straightforward, minor modification. This is required by the fact that we have no discrete structure in time direction. Formally, it looks like time remaining continuous. This requires a mixed form for the definition of the gauge field: The temporal component $A_0(n,t) \in \mathfrak{g}$ is, like in the continuous case, a function with values in the Lie algebra $\mathfrak{g}$, and defined on the lattice nodes. Instead, the spatial (vector potential) part $A_i$ is described, as usual for Wilson gauge fields, by Lie group valued functions $U(n,i,t)\in G$ located on the edges $n,n+h_i$ of the lattice. The most important, defining property of the Wilson gauge field remains unchanged too: The lattice gauge symmetry is defined by a gauge-group-valued lattice function $g(n): \Z^3\to G$, which acts pointwise on the lattice \CAZ\/ and is uniquely defined by a gauge action $G\times\C\times\A\to\C\times\A$. The gauge transformation acts in the following way:
\begin{subequations}\label{def:WilsonAction}
\begin{eqnarray}
\psi^i_\mu(n,t) &\to& (g(n,t)\psi)^i_\mu(n,t), \\
A_0(n,t)  &\to& g(n,t)A_0(n,t)g^{-1}(n,t) - (\pd_t g(n,t))g^{-1}(n,t),\\
U(n,i,t)  &\to& g(n,t)U(n,i,t)g^{-1}(n+\hi,t).
\end{eqnarray}
\end{subequations}
This definition of the gauge action \eqref{def:WilsonAction} shows that not all imaginable gauge actions may be defined in this way. Indeed, the gauge action can act only on the generation and color indices. Inside a doublet \CL, it can act only in a very restricted way: An electroweak doublet with fixed generation $i$ and color $\mu$ is represented on the lattice as a lattice field \CZ, so that there is only a single complex number $\psi^i_\mu(n)$ in each lattice node. The only possible Wilson gauge action on the lattice \CZ\/ is, obviously, an action of $U(1)$, which means, that we obtain the same charge on all parts of the doublet.

Now, this already allows to compute the maximal possible Wilson gauge action, which is compatible with our postulates \ref{postulate:symplectic} and \ref{postulate:Euclidean}. It should be an unitary group. It acts on all generations in the same way, and preserves the generations, thus, does not act on the generation index $i$. Then, it acts with the same charge on all parts of electroweak doublets, thus, cannot act on the doublet indices $\k$. Thus, it can act only on the remaining index $\mu$. This gives $U(4)$ as the maximal gauge group. Moreover, to commute with translations, it has to leave the translational direction \c\/ in the leptonic sector invariant. But, because it has the same charge on all parts of the leptonic doublets, it has to act trivially on the whole leptonic sector $\mu=0$. What remains is the group $U(3)$ acting on the color index $\mu>0$. Its special subgroup $SU(3)$ can be, obviously, identified with the color group $SU(3)_c$ of the SM. What remains is the diagonal $U(1)_B$ with the baryon charge $I_B$.

Having found an upper bound, let's consider the question if these Wilson gauge fields will appear, in some natural way, in our condensed matter model. For this purpose, let's assume that there is some other material between the cells. Inhomogeneities of this material will influence the cells, thus, lead to some modification of the equation for the cells. In some approximation, the material located between two neighbour cells will influence only those parts of the equation which connect these two cells. Then, a Wilson gauge field corresponds to such an influence which may be compensated by a modification of the state and momentum of one of the neighbour cells. It seems reasonable to expect that such influences of the material between the cells appear. 

Thus, the consideration of Wilson lattice gauge fields has given us, almost exactly, an important part of the SM gauge group --- the strong interactions.

\subsection{Correction terms for lattice deformations}\label{weak}

While the consideration of Wilson gauge fields is a sufficiently trivial modification of standard Wilson gauge fields, the incorporation of weak interactions requires a non-standard approach to lattice gauge theory. This new approach is far away from being completed. Nonetheless, it already gives a nice correspondence between the properties of the gauge groups which may be, in principle, obtained in this way, and the gauge group of the SM.

Assume our lattice $\Z^3$ is not exactly regular but slightly deformed. This requires also a modification of the lattice Dirac equation. What can be said about the general form of the corresponding correction terms?

First, a deformation of the lattice is certainly no reason to use different lattice equations for the different components $\psi^i_\mu(n)\in\CAZ$. However deformed the lattice, the correction coefficients are of geometric nature: They depend only on the geometry of the deformed lattice. Thus, the deformed lattice equation will be an equation on the same bundle \CL, independent of the generation and color indices $i$ and $\mu$. This leads to the following
\begin{thesis}\label{thesis:weak}
Correction terms for lattice deformations preserve doublets \CL\/ and act on all doublets in the same way.
\end{thesis}
But this is a signature of weak forces. Thus, it seems natural to postulate a connection between weak interactions and correction terms for lattice deformations.

Let's consider, therefore, possible correction terms in more detail. We start with a regular lattice \CZ, with a function $\psi(n)$ on it, and an undistorted lattice Dirac equation. For a slightly deformed lattice, the value in a regular lattice node $x^i=m_ih$ is no longer $\psi(m)$, but has to be interpolated using all neighbour nodes. Thus, we correct now every occurrence of $\psi(m)$ by a weighted sum over values $\psi(m+\h)$ on neighbour nodes (including the node $m$ itself):
\begin{equation}
 \psi(m) \to \psi(m) + \sum_\h \hat{A}^\h_p(n)\psi(m+\h)
\end{equation} 
with some set of geometric coefficients $\hat{A}^\h_p(n)$. These coefficients depend, in general, on the basic node $n$ of the lattice equation containing a term with $\psi(m)$, on the direction of the neighbour $\h\in\Z^3$, and, moreover, on the occurrence $p$ of the term $\psi(m)$ in the undistorted Dirac equation for node $n$. Using the lattice shift operator $\tau_\h: \psi(m)\to\psi(m+\h)$, we can rewrite the expression as
\begin{equation}
 \psi(m) \to \Bigl(1 + \sum_\h \hat{A}^\h_p(n)\tau_\h\Bigr)\; \psi(m).
\end{equation} 
Now, instead of the lattice shift operators $\tau_\h$, which do not commute with the Dirac equation, we prefer to use another set of operators associated with lattice shifts, namely the operators $\tilde{\tau}_\h\in\mathcal{A}$ of theorem \ref{th:shiftAlgebra}, which commute with the Dirac equation. Fortunately, this is possible. It requires only a redefinition of the coefficients $\hat{A}^\h_p(n) \to \tilde{A}^\h_p(n)$: We can replace the $\tau_\h$ by $\tilde{\tau}_\h = o^\h(n)\tau_\h$, with $o^\h(n)\in\C$, and put the coefficients $o^\h(n)$ into the $\hat{A}^\h_p(n)$ with $\tilde{A}^\h_p(n) = \hat{A}^\h_p(n)(o^\h(n))^{-1}$. Using these pseudo-shift operators $\tilde{\tau}_\h\in\mathcal{A}$, together with the geometric coefficients $\tilde{A}^\h_p(n)$, gives
\begin{equation}
 \psi(m) \to \Bigl(1 + \sum_\h \tilde{A}^\h_p(n)\tilde{\tau}_\h\Bigr)\; \psi(m).
\end{equation} 
We have seven occurrences $p\in\{0,i\pm\}$ of $\psi$ in the lattice Dirac equation of node $n$, which gives
\begin{equation}\label{eq:DiracLatticeDeformed}
\begin{split}
	i\pd_t \bigl(\delta^{\k'}_{\k}+\tilde{A}^\h_{0 }(n)(\tilde{\tau}_\h)^{\k'}_{\k}\bigr)\psi_{\k'}(n) =  -i(\alpha^i)_\k^{\k'}
	\Bigl(& \bigl(\delta^{\k''}_{\k'}+\tilde{A}^\h_{i+}(n)(\tilde{\tau}_\h)^{\k''}_{\k'}\bigr)\psi_{\k''}({n+h_i})\\
	-&\bigl(\delta^{\k''}_{\k'}+\tilde{A}^\h_{i-}(n)(\tilde{\tau}_\h)^{\k''}_{\k'}\bigr)\psi_{\k''}({n-h_i})\Bigr).
\end{split}
\end{equation}
Introducing the denotations
\begin{align}
A_i^\h(n) &= -i(\tilde{A}^\h_{i+}(n)-\tilde{A}^\h_{i-}(n)), & A_0^\h(n) &= -i\pd_t \tilde{A}^\h_0(n),\\
\breve{A}_i^\h(n) &= -i\frac{1}{2}(\tilde{A}^\h_{i+}(n)+\tilde{A}^\h_{i-}(n)), &\breve{A}_0^\h(n) &= -i\tilde{A}^\h_0(n),
\end{align}
this becomes
\begin{equation}\label{eq:DiracLatticeWeak}
\begin{split}
	i\bigl(\delta^{\k'}_{\k}&+i\breve{A}_0^\h(n)(\tilde{\tau}_\h)^{\k'}_{\k}\bigr)\pd_t\psi_{\k'}(n)=\\
	&(\alpha^i)_\k^{\k'}\Bigl(-i
		\bigl(\delta^{\k''}_{\k'}+\breve{A}^\h_{i}(n)(\tilde{\tau}_\h)^{\k''}_{\k'}\bigr)
		\bigl(\psi_{\k''}({n+h_i})-\psi_{\k''}({n-h_i})\bigr)\\
		&\phantom{(\alpha^i)_\k^{\k'}(} + A_i^\h(n) (\tilde{\tau}_\h)^{\k''}_{\k'}
			 \frac{\psi_{\k''}({n+h_i})+\psi_{\k''}({n-h_i})}{2}\Bigr)\\
		&+ A_0^\h(n) (\tilde{\tau}_\h)^{\k'}_{\k} \psi_{\k'}(n)
\end{split}
\end{equation}
Now, there are two straightforward simplifications, which we can use, once we are interested only in the large distance limit. First, while $A_i^\h(n)$ interacts with terms which, in the large distance limit, become $\psi_\k(x)$, the $\breve{A}_i^\h(n)$ interact with their derivatives $\pd_\mu\psi_\k(x)$. We leave only the lowest order interaction terms, omitting the interaction terms containing the derivatives, therefore, the terms containing $\breve{A}_\mu^\h(x)$. Then, instead of a summation over all possible neighbours $\h\in\Z^3$, we can restrict the summation to the eight ``non-trivial'' basis lattice shifts $\hb\in\{0,1\}^3$: Terms which differ only by even lattice shifts become almost identical. This gives
\begin{equation}\label{eq:gaugelike}
\begin{split}
	i\pd_t  \psi_{\k}(n) = & (\alpha^i)_\k^{\k'}
		\left(-i(\psi_{\k'}({n+h_i})-\psi_{\k'}({n-h_i}))\phantom{\frac{\psi_{\k''}}{2}}\right.\\
		&\phantom{=  (\alpha^i)_\k^{\k'}(-}\left.
			+ A_i^\hb(n) (\tilde{\tau}_\hb)^{\k''}_{\k'} \frac{\psi_{\k''}({n+h_i})+\psi_{\k''}({n-h_i})}{2}\right)\\
		&+ A_0^\hb(n) (\tilde{\tau}_\hb)^{\k'}_{\k} \psi_{\k'}(n).
\end{split}
\end{equation}
Remember now that the operators $\tilde{\tau}_\hb$ are lattice approximations of the set of operators $\{1,\g^5,2I_i,2I_i\g^5\}$ on the staggered lattice (see theorem \ref{th:shiftAlgebra}), and that these operators generate the Lie algebra of $U(2)_L\times U(2)_R$ --- the left-right-symmetric extension of the weak gauge group $SU(2)_L$. Then, compare \eqref{eq:gaugelike} with the continuous Dirac equation which interacts with some gauge field $A_\mu^\hb(x)$ via some representation $\hat{\tau}_\hb\in\{1,\g^5,2I_i,2I_i\g^5\}$:
\begin{equation}
	i\pd_t 	\psi_{\k}(x) =  (\alpha^i)_\k^{\k'}
		(-i\pd_i + A_i^\hb(x)\hat{\tau}_\hb) \psi(x)
		+ A_0^\hb(x) \hat{\tau}_\hb \psi(x).
		\label{eq:DiracWeakGauge}
\end{equation}
We see that (\ref{eq:gaugelike}) is simply a discretization of (\ref{eq:DiracWeakGauge}), which correctly takes into account that we have a staggered lattice. Remembering the lecture of fermion doubling, we should avoid premature claims that the continuous limit of \eqref{eq:DiracLatticeDeformed} is \eqref{eq:DiracWeakGauge}. Nonetheless, for a given lattice equation \eqref{eq:DiracLatticeDeformed}, we can compute low energy effective fields  $A_\mu^\hb(x)$ by averaging over the $A_\mu^\hb(n)$. And these effective fields $A_\mu^\hb(x)$ interact with fermions like gauge fields with the gauge group $U(2)_L\times U(2)_R$. This can be summarized in the following theorem:

\begin{theorem}\label{th:weakGaugeField}
A lattice deformation described by \eqref{eq:DiracLatticeDeformed} defines effective low energy fields $A_\mu^\hb(x)$. These fields interact with fermions in the same way as the gauge field of the left-right-symmetric extension $U(2)_L\times U(2)_R$ of the weak gauge group $SU(2)_L$.
\end{theorem}

Note here that it does not follow that for each continuous field configuration $A^\hb_\mu(x)$ exists a physically meaningful deformed lattice giving this field in the continuous limit. Instead, we obtain, below, two additional restrictions for the $A^\hb_\mu(x)$. In principle, we have not even proven here that there exist physically meaningful deformed lattices which give nontrivial fields $A_\mu^\hb(x)$ at all. Despite this, the construction of the $A_\mu^\hb(x)$ is not useless, but shifts the burden of argumentation: It is, now, the thesis that the fields $A_\mu^\hb(x)$ do not give the full group $U(2)_L\times U(2)_R$, which requires justification.

Note also that we do not claim here the existence of some effective gauge symmetry. What we have found, is only a description in terms of effective fields $A^\hb_\mu(x)$, which interact with the fermions in the same way as gauge fields. It is only the continuous limit of the interaction part which shows some gauge invariance. Different gauge-equivalent effective fields $A^\hb_\mu(x)$ describe different geometric coefficients $\hat{A}^\h_p(n)$, thus, different deformed lattices. That means, on the fundamental level we have no gauge symmetry. Moreover, we have no theory about the equations of motion or a Lagrange formalism for these effective fields, nor on the fundamental level, nor in the continuous limit. This has to be left to future research.

This theorem also allows the existence of other effective fields, which interact differently with the fermions. Here we have in mind the continuous limit of the fields $\breve{A}_i^\h(n)$, which give something like fields $\breve{A}_\mu^\hb(x)$ interacting with the derivatives of $\psi$. Moreover, at the current state of research we cannot exclude, as well, that something similar to fermion doubling happens too, giving some other effective fields in the continuous limit. Nonetheless, the lattice equation (\ref{eq:DiracLatticeDeformed}) is already sufficient to restrict the interaction between these possible other fields and the fermions: Whatever these fields, the interaction is restricted to the maximal group $U(2)_L\times U(2)_R$:

\begin{theorem} \label{th:weakInteraction} For lattice deformation described by the lattice equation \eqref{eq:DiracLatticeDeformed}, the resulting low-energy effective fields can interact with the fermions only via the set of operators $\hat{\tau}_\hb \in \{1,\g^5,2I_i,2I_i\g^5\}$, which generate the Lie algebra representation of the group $U(2)_L\times U(2)_R$.
\end{theorem}

Indeed, the lattice coefficients  $\tilde{A}^\h_p(n)$ interact with the fermion field $\psi$ only via the operators $\tilde{\tau}_\h$. But these operators become, in the continuous limit, $\hat{\tau}_\hb \in \{1,\g^5,2I_i,2I_i\g^5\}$. And interaction with these generators cannot give more than an action of the group $U(2)_L\times U(2)_R$.

Now, it is reasonable to assume that the postulates \ref{postulate:symplectic}, \ref{postulate:Euclidean} hold for deformed lattices as well as for undeformed ones. Thus, we apply these postulates to our gauge-like correction terms as well. This leads to additional restrictions. While the preservation of the symplectic structure and rotational symmetry does not give anything new ($U(2)_L\times U(2)_R$ is already unitary, preserves generations, and acts on all generations in the same way, thus, already preserves the symplectic structure as well as rotational symmetry), translational symmetry requires a further reduction of the maximal possible gauge group. Indeed, there has to be a preserved translational direction \c\/ in the leptonic sector. But the maximal group $U(2)_L\times U(2)_R$ does not leave any direction invariant. Thus, independent of the choice of the direction \c, it cannot commute with translations. Any group, which commutes with translations, should be a nontrivial subgroup of $U(2)_L\times U(2)_R$, and leave at least one particular direction \c\/ invariant.

There are several possibilities for the translational direction \c: It may be left-handed, right-handed, or none of the above.  The last case gives a more rigorous restriction of the group: The left-handed part $\frac{1-\g^5}{2}\c$ as well as the right-handed part $\frac{1+\g^5}{2}\c$ would have to be preserved separately. Thus, to compute the maximal possible gauge group, we can ignore the last case.

If the translational direction \c\/ is right-handed, we have the group $U(2)_L$ preserving this direction, while $U(2)_R$ has to be reduced to some subgroup $U(1)_R\subset U(2)_R$. For this group, the charge of the translational direction \c\/ should be $0$. Without restriction of generality, this charge can be taken as $I_R = \frac{1+\g^5}{2}(I_3-\frac{1}{2})$, so that we obtain $U(2)_L\times U(1)_R$ as the maximal possible gauge group. Similarly, for left-handed \c, we obtain the equivalent maximal gauge group $U(2)_R\times U(1)_L$.

\begin{theorem}\label{th:weakGroup}
The maximal gauge group, which may be obtained from correction terms for lattice deformations, and is compatible with \E\/ symmetry and symplectic structure, is a subgroup $U(2)\times U(1)$ of $U(2)_L\times U(2)_R$. Without restriction of generality, this maximal group may be identified with $U(2)_L\times U(1)_R$, generated by the left-handed chiral $U(2)_L$, and the right-handed chiral $U(1)_R$ with charge $I_R = \frac{1+\g^5}{2}(I_3-\frac{1}{2})$.
\end{theorem}

\subsection{The EM field}

At a first look, the EM field does not fit into any of the two classes of gauge-like lattice fields: It acts nontrivially inside doublets, thus, is not a Wilson field. On the other hand, it has different charges on leptons and quarks, thus, does not fit into the our scheme for correction terms for lattice deformations.

Nonetheless, we have already obtained it: It appears as a linear combination of the two types of gauge-like fields we have obtained. For the EM field we have
\begin{equation}
 U(1)_{em} \subset S(U(1)_B\times U(2)_L\times U(1)_R).
\end{equation} 
Indeed, the EM charge
\begin{equation}
 Q = 2I_B + (I_3-\frac{1}{2}) = 2I_B + \frac{1-\g^5}{2}(I_3-\frac{1}{2}) + \frac{1+\g^5}{2}(I_3-\frac{1}{2}).
\end{equation}
is a combination of different charges: $I_B$ comes from the diagonal of $U(3)_c$ of Wilson gauge fields, $\frac{1-\g^5}{2}(I_3-\frac{1}{2})$ is part of $U(2)_L$ of weak interaction, and $\frac{1+\g^5}{2}(I_3-\frac{1}{2})$ is the charge of $U(1)_R$. A similar decomposition exists for the hypercharge:
\begin{equation}\label{eq:Ydef}
 Y = 2I_B - \frac{1-\g^5}{4} + \frac{1+\g^5}{2}(I_3-\frac{1}{2}).
\end{equation}

\subsection{The main result}

Thus, all gauge fields of the SM are part of the maximal possible gauge group which can be constructed with our two ways to put gauge fields on the lattice:
\begin{equation}
U(3)_c \times U(2)_L \times U(1)_R \supset G_{SM} \cong SU(3)_c \times SU(2)_L \times U(1)_Y.
\end{equation}
Once these two types of lattice gauge fields are sufficient, let's add, in agreement with Ockham's razor, another postulate:
\begin{postulate} \label{postulate:lattice}
All gauge-like fields can be obtained as effective fields from the following two types of lattice fields:
\begin{itemize}
 \item Wilson lattice gauge fields;
 \item Correction terms for lattice deformations.
\end{itemize}
\end{postulate}

With this postulate, the maximal possible gauge group is (without restriction of generality) $U(3)_c \times U(2)_L \times U(1)_R$. It contains only three fields which are not part of the SM:  The diagonal of $U(3)_c$, the diagonal of $U(2)_L$, and $U(1)_R$. But, according to \eqref{eq:Ydef}, one linear combination of all three gives the hypercharge group $U(1)_Y$. Thus, we have only two additional gauge fields in the maximal possible gauge group.

A further reduction can be obtained with the following postulate:
\begin{postulate} \label{postulate:special}
The gauge group should be a special group.
\end{postulate}
This postulate can be justified using general properties of effective forces in condensed matter theory.  Volovik \cite{Volovik} writes: ``An equilibrium homogeneous ground state of condensed matter has zero charge density, if charges interact via long range forces. For example, electroneutrality is the necessary property of bulk metals and superconductors; otherwise the vacuum energy of the system diverges faster than its volume. \ldots The same argument can be applied to `Planck condensed matter', and seems to work. The density of the electric charge of the Dirac sea is zero due to exact cancellation of electric charges of electrons, $q_e$, and quarks, $q_u$ and $q_d$, in the fermionic vacuum: $Q_{vac} = \sum_{E<0} (q_e+3q_u+3q_d) = 0$.''
\footnote{
In a similar way, a condensed matter approach promises to solve the cosmological constant problem \cite{Volovik}: ``The $^3$He analogy suggests  that a zero value of the cosmological term in the equilibrium vacuum is dictated by the Planckian or trans-Planckian degrees of freedom:  $\partial S_{\mathrm{vac}}/\partial g_{\mu\nu}=0$ is the thermodynamic equilibrium condition for the `Planck condensed matter'. Thus the equilibrium homogeneous vacuum does not gravitate. Deviations of the vacuum from its equilibrium can gravitate.''
}
Thus, to obtain neutrality of the ground state, the trace of the charges has to be zero. So the group has to be a special group, with determinant $1$. This postulate allows to get rid of the field $U(1)_B$. This seems necessary, because already a very coarse consideration suggests that an $U(1)_B$ gauge field should be at least strongly suppressed: The Earth would be heavily charged, leading to a repulsive force on all baryonic matter. Thus, it should be weaker than gravity. A combination of gravity with this force would, moreover, lead to a violation of the EEP, thus, can be distinguished from pure gravity by tests of the EEP. Thus, an $U(1)_B$ field undetected by existing EEP tests should be much weaker than gravity, and, therefore, extremely suppressed in comparison with the EM field.

Thus, there remains only a single additional gauge field. It can be chosen as an ``upper axial gauge field'' $U(1)_{U}$, which acts on all upper particles (upper quarks and leptons) axially, with charge
\begin{equation}
I_{U} = \g^5(I_3-\frac{1}{2}). 
\end{equation}
There is also a hyper-variant $U(1)_{\hat{U}}$, which commutes with all SM gauge fields. It's charge is
\begin{equation}
I_{\hat{U}} = I_{U} + \frac{1-\g^5}{2}I_3 = I_R + \frac{1-\g^5}{4} = \frac{1+\g^5}{2}I_3 - \frac{\g^5}{2}.
\end{equation}

This can be summarized in the following
\begin{theorem}\label{th:main}
The maximal possible gauge group $G_{max}$ with the following properties:
\begin{itemize}
\item Postulate \ref{postulate:symplectic}: preservation of the symplectic structure;
\item Postulate \ref{postulate:Euclidean}: preservation of \E\/ symmetry;
\item Postulate \ref{postulate:lattice}: realization on the lattice as
\begin{itemize}
\item Wilson gauge fields;
\item Correction terms for lattice deformations;
\end{itemize}
\item Postulate \ref{postulate:special}: restriction to special group;
\end{itemize}
is $S(U(3)_c \times U(2)_L \times U(1)_R) \cong G_{SM} \times U(1)_{\hat{U}}$.

It acts, without restriction of generality, on the fermions with the standard action of $G_{SM}$ and the $U(1)_{\hat{U}}$ charge
\begin{equation}
 I_{\hat{U}} = \frac{1+\g^5}{2}I_3 - \frac{\g^5}{2}.
\end{equation}
\end{theorem}

Now, the action of $G_{max}$ is anomalous. Thus, $G_{SM}$ may be characterized as a maximal anomaly-free subgroup of $G_{max}$. To obtain the SM, we could, as well, add anomaly freedom as an additional postulate:
\begin{postulate}\label{postulate:noanomaly}
 The action of the group has to be anomaly-free.
\end{postulate}

Because $U(1)_B$ defines also an anomalous extension of $G_{SM}$, we can, in this case, even omit postulate \ref{postulate:special}, and obtain the following theorem:
\begin{theorem}\label{th:main1}
The SM gauge group $G_{SM}$, with it's standard action, is a maximal group with the following properties:
\begin{itemize}
\item Postulate \ref{postulate:symplectic}: preservation of the symplectic structure;
\item Postulate \ref{postulate:Euclidean}: preservation of \E\/ symmetry;
\item Postulate \ref{postulate:lattice}: realization on the lattice as
\begin{itemize}
\item Wilson gauge fields;
\item Correction terms for lattice deformations;
\end{itemize}
\item Postulate \ref{postulate:noanomaly}: anomaly freedom.
\end{itemize}
\end{theorem}

But note that the standard rejection of gauge groups with anomalies is based on the canonical BRST approach to gauge field quantization, which requires exact gauge invariance. Instead, in our approach, gauge degrees of freedom are physical, and the Hilbert space is definite from the start, as discussed below in section \ref{sec:quantumGaugeFields}. Therefore, the standard argumentation against anomalous gauge fields fails: Without an indefinite Hilbert space structure, and without factorization of the (now physical) gauge degrees of freedom, there is no way to obtain a non-unitary theory. Therefore it seems premature to reject the extension $U(1)_{U}$ only because of the resulting gauge anomaly. So the author prefers, at the current state, theorem \ref{th:main}, and hopes for future observation of a new gauge field $U(1)_{U}$.

On the other hand, it seems also premature to claim that our approach predicts the additional field $U(1)_{U}$. Indeed, we have to expect, that the anomaly leads to nontrivial effects during renormalization. As a result, the field $U(1)_{U}$ may as well become effectively suppressed.

But even if such effects of renormalization suppress the anomalies, it would be unclear which of the non-anomalous subgroups survives. In itself, the baryon charge $U(1)_B$ is, as a vector gauge field, non-anomalous. So, why do we have the SM gauge group instead of the maximal non-anomalous extension of $U(1)_B$? 

Postulate \ref{postulate:special} would give an answer. And, therefore, it seems a good idea to preserve it. 

\section{Fermion quantization} \label{fermionQuantization}

Our approach to fermion quantization is a drastic departure from the common wisdom of quantum field theory. According to the standard approach, fermions do not have a classical configuration in the usual sense. Following Berezin \cite{Berezin}, instead of functions on a classical configuration space, we have, as some sort of classical limit, only Grassmann-valued fields.

This is, obviously, incompatible with our geometric interpretation of fermion doublets as \CL\/ or the lattice model \CZ, which are classical, commuting, real fields, with a standard symplectic structure on the phase space. The appropriate way to quantize them would be canonical quantization. This seems, at a first look, impossible --- last but not least, we obtain in such a way only commuting operators.

Despite this, we present here a way to obtain anticommuting fermion fields via canonical quantization. It consists of two parts, with a canonically quantized \B-valued field (spin field) as the intermediate step. The first part -- to obtain a \B-valued field from an \R-valued field -- is rather straightforward: All we need is a \B-degenerated potential $V(\varphi)$. The lowest energy states, then,  define a \B-valued field theory.

The non-trivial step is from spin fields to fermion fields, or from commuting to anticommuting operators. Surprisingly, the lattice operator algebras appear to be isomorph. The isomorphism is, essentially, known in Clifford algebra theory. But this isomorphism is, first, nonlocal, and, second, not natural. It depends on some ordering between different lattice nodes. Moreover, it requires, for $dim>1$, a nontrivial approximation of the Hamilton operator. Despite these caveats, the isomorphism between lattice fermion operators and \B-valued ``spin field'' operators proves that fermions may be obtained via canonical quantization of real fields.

In this paper, we need only canonical quantization of lattice theories with configuration space $Q=\R(\Z^3)$ resp. $Q=\B(\Z^3)$. But our considerations here do not depend on the dimension $d=3$, so we consider here the more general case $Q=\R(\Z^d)$ resp. $Q=\B(\Z^d)$. Canonical quantization consists of the definition of operators on the Hilbert space $\mathscr{L}^2(Q)$, and a Schr\"{o}dinger equation
\begin{equation}
	i\pd_t \Psi(q,t)=H\Psi(q,t), \qquad q \in Q=\mathscr{F}(\Z^d,Y),\qquad t \in \R.
	\label{eq:Schroedinger}
\end{equation}
Thus, we always have continuous time. Note that in our condensed matter interpretation the lattice $\Z^3$ is not a ``discretization of space'' $\R^3$ itself. Instead, it enumerates elementary cells located in a continuous $\R^3$, where the state of each cell is described by some affine transformation $\A\subset \R^{12}$ of $\R^3$. Nonetheless, in this section, there is no difference between our lattice $\Z^3$ and a ``discretization of space'' $\Z^3\subset \R^3$, and our geometric interpretation of $\R^{12}$ in terms of \A\/ plays no role.

\subsection{From spin fields to fermion fields}\label{fermion2spin}

Spin fields have the configuration space $\B(\Z^d)$. On each lattice node $n\in \Z^d$ we have the Pauli matrices $\sigma^i_n$ as operators:
\begin{equation}
\sigma_n^i\sigma_n^j = \delta_{ij}+i\varepsilon_{ijk}\sigma_n^k.\;
\label{eq:sigma:product}
\end{equation}
Spin field operators on different nodes commute:
\begin{equation}
\left[\sigma^i_m, \sigma^j_n\right]  = 2i\delta_{mn}      \varepsilon_{ijk}\sigma^k_n.
\label{eq:sigmaCommutation}
\end{equation}

Instead, following Berezin\cite{Berezin}, the fermion field operators $\psi_n, \psi^*_n$ are usually considered to be of completely different nature. They do not fit into the canonical quantization scheme. Especially there is no configuration space $Q$. Operators related to different nodes do not commute. Instead, they anticommute:
\begin{equation}
\{\psi_m,\psi^*_n\} = \delta_{mn},\;\{\psi^*_m,\psi^*_n\}=\{\psi_m,\psi_n\}=0.
\label{eq:psiAnticommutation}
\end{equation}
This difference seems to forbid an identification of fermions with spin fields. Despite this, the two operator algebras appear to be isomorph:

\begin{theorem} The operator algebra of spin field operators \eqref{eq:sigma:product}, \eqref{eq:sigmaCommutation} and the algebra of fermion operators \eqref{eq:psiAnticommutation} are isomorph. The isomorphism is not natural, but depends on the choice of an order $>$ between the lattice nodes.
\end{theorem}

To prove this, let's at first transform the operator algebras in each node into an equivalent form, by defining operators $\psi^i_n$:
\begin{equation}
\psi_n^1 = \psi_n + \psi_n^*,\; \psi_n^2 = -i(\psi_n - \psi_n^*),\;\psi_n^3 = -i\psi_n^1\psi_n^2.
\label{eq:def:psi^i}
\end{equation}
This gives
\begin{equation}
\psi_n^i\psi_n^j = \delta_{ij}+i\varepsilon_{ijk}\psi_n^k,\;
\label{eq:psi:product}
\end{equation}
similar to (\ref{eq:sigma:product}). Now, for a given order $>$ between the nodes of the lattice, the isomorphism is defined by the following formulas:
\begin{subequations}\label{eq:psisigmaconversion}
\begin{align}
\label{eq:psi2sigma}
\psi^{1/2}_n &= \sigma^{1/2}_n \prod_{m>n}{\sigma^3_m},& \psi^3_n&=\sigma^3_n,\\
\label{eq:sigma2psi}
\sigma^{1/2}_n &= \psi^{1/2}_n \prod_{m>n}{\psi^3_m},& \sigma^3_n&=\psi^3_n.
\end{align} 
\end{subequations}

Let's note that this isomorphism is well-known in the theory of Clifford algebras and allows to establish the isomorphism
\begin{equation}
\textit{Cl}^{N,N}(\R)\cong M_2(\textit{Cl}^{N-1,N-1}(\R))\cong M_{2^N}(\R).
\label{eq:Clifford}
\end{equation}
Indeed, the operators $\psi^1_n$ and $i\psi^2_n$ generate (for a finite lattice with N nodes) the Clifford algebra $\textit{Cl}^{N,N}(\R)$. On the other hand, $M_{2^N}(\R)$ is the operator algebra on the $2^N$-dimensional space of \B-valued functions on the same lattice.

Note also that (different from the $\sigma^i_n$) the operators $\psi_n^i$ do not act as local operators on the lattice. Instead, they act like $\sigma^3_m$ on other nodes $m>n$. This is a necessary property of such an isomorphism, because, obviously, any local combination of the commuting local operators $\sigma^i_n$ can give only to another set of commuting local operators.

As a consequence, a Hamilton operator which ``looks local'' in terms of the $\psi^i_n$ may appear nonlocal in terms of the $\sigma^i_n$ (which we consider to be ``truly local'' operators) and reverse. Fortunately, there are important examples of operators where this does not happen. First, there is the operator
\begin{equation}\label{eq:def:H0}
  H_0\  =	- \frac{1}{2}\sum_n{\sigma_n^3} =\  \frac{1}{2}\sum_n \psi^*_n\psi_n-\psi_n\psi^*_n
\end{equation}

Let's consider now operators with interactions between neighbour nodes. We are (for reasons which become obvious later) especially interested in the following linear combination:
\begin{equation}
	\label{eq:def:HDd}
	H_D\  = \frac{1}{2}\sum_{n,i}\sigma^1_n\sigma^1_{n+h_i}-\sigma^2_n\sigma^2_{n+h_i}
\end{equation}
where $h_i$ are the $d$ basic lattice shifts in the d-dimensional lattice $\Z^d$.

In the one-dimensional case, we have a natural (up to the sign) order $>$. For this order, we obtain:
\begin{equation}
	\label{eq:def:HD1}
	H^{(1)}_D\  =	i\frac{1}{2}\sum_n(\psi^1_n\psi^2_{n+1}+\psi^2_n\psi^1_{n+1})\ =
	\ \sum_n \psi_n\psi_{n+1}-\psi^*_n\psi^*_{n+1}
\end{equation}
Note that our operator $H^{(1)}_D$ itself is symmetric for spatial inversion $n\to -n$. But the representation in the asymmetric (in terms of the $\sigma^i_n$) operators $\psi^i_n$ hides this symmetry.

\subsection{The case of higher dimensions}

Unfortunately, the transformation of the Hamilton operator in higher dimensions is not that simple. What we can obtain is only an approximation
\begin{equation}
	\label{eq:def:tHDd}
	H^{(d)}_D\ \approx\  \tilde{H}^{(d)}_D\  =
	\sum_{n,i} \alpha^n_{n+h_i} (\psi_n\psi_{n+h_i}-\psi^*_n\psi^*_{n+h_i})
\end{equation}
where
\begin{equation}
	\label{eq:alphadef}
	\alpha^n_{n+h_i} = \left\{ \begin{aligned}
		1 &\quad \text{if}\quad  n < n+h_i\\
		-1 &\quad \text{else}\end{aligned} 
	\right. \
\end{equation}
The accuracy of this approximation obviously depends on the order $>$.  Indeed, the error
\begin{equation}
	\sigma^1_n \sigma^1_{n'} \approx \psi^1_n \psi^2_{n'} =  \sigma^1_n \sigma^1_{n'}\prod_{n<m<n'}{\sigma^3_m},
	\label{eq:prodnonlocal}
\end{equation}
resp. for $\sigma^2_n \sigma^2_{n'}$, depends on the number and location of the nodes $m$ located ``between'' (according to the chosen ordering) the ``neighbour'' (according to the lattice $\Z^d$) nodes $n,n'$. Now, instead of the simple lexicographic order (which gives $\alpha^n_{n+h_i}=1$) we propose to use another, more sophisticated order we name ``alternating lexicographic order''.

It has to be acknowledged that this order has been designed to give the result below. Fortunately, we can justify this choice of an order in another way: It gives a better approximation of the original Hamiltonian operator: Some algebraic properties of the original terms may be preserved exactly.

Note that our interaction terms can be represented as a function of the differences of the operator $\sigma^1_n$ and its shift:
\begin{equation}
	\sigma^1_n\sigma^1_{n+h_i} = 1 - \frac{1}{2}((1-\tau_i)\sigma^1_n)^2,
	\label{eq:differenceproperty}
\end{equation}
where $\tau_i$ is the shift operator on the lattice. This follows from $(\sigma^1_n)^2=1$ and the commutation relation $[\sigma^1_n,\tau_i\sigma^1_n]=0$. Now, we propose to use an order which allows to preserve these properties exactly. That means, we want to replace the $\sigma^1_n$ by some $\tilde{\sigma}^1_n$ with exactly the same properties:
\begin{equation}
	(\tilde{\sigma}^1_n)^2=1,\;  [\tilde{\sigma}^1_n,\tau_{i}\tilde{\sigma}^1_{n}]=0,
	\label{eq:approximationalproperties}
\end{equation}
so that
\begin{equation}
	\sigma^1_n\sigma^1_{n+h_i} \approx \tilde{\sigma}^1_n\tilde{\sigma}^1_{n+h_i}= 1 -
	\frac{1}{2}((1-\tau_i)\tilde{\sigma}^1_n)^2.
\end{equation}

\begin{figure}
\includegraphics[angle=0,width=0.8\textwidth]{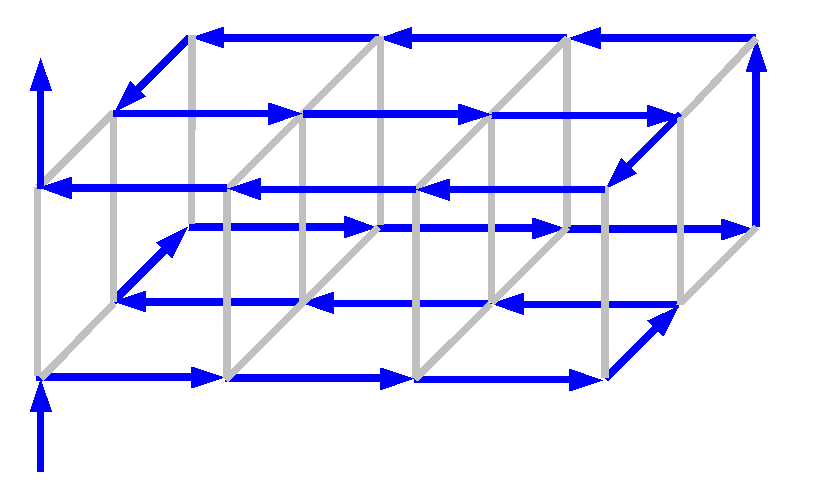}
\caption{\label{fig:order}The alternating lexicographic order}
\end{figure}

For the simple lexicographic order, we have no way to define such $\tilde{\sigma}^i_{n}$. But it is possible for the alternating lexicogrpahic order. Let's define this order by induction. Let $>_k$ be the order defined for a k-dimensional lattice $\Z^k$, and $\pi_k$ the projection on this lattice defined by the first $k$ coordinates.  Then we define $>_{k+1}$ by the following properties:
\begin{itemize}
	\item if $n_{k+1} \lessgtr m_{k+1}$ then $n\lessgtr_{k+1}m$;
	\item else if $n_{k+1} (= m_{k+1})$ is even and $\pi_k n \lessgtr_{k} \pi_k m$ then $ n \lessgtr_{k+1} m$;
	\item else if $n_{k+1} (= m_{k+1})$ is odd and $\pi_k n \lessgtr_{k} \pi_k m$ then $ n \gtrless_{k+1} m$.
\end{itemize}
Thus, we use the inverse order inside the odd planes. Now the interaction term can be splitted in the following way:
\begin{equation}
	\sigma^{1/2}_n \sigma^{1/2}_{n+h_i}\prod_{n<m<n+h_i}{\sigma^3_m} =
	\tilde{\sigma}^{1/2}_n \tilde{\sigma}^{1/2}_{n+h_i}
	\label{eq:proddecomposition}
\end{equation}
with
\begin{equation}
	\tilde{\sigma}^{1/2}_n=\sigma^{1/2}_n
		\prod_{\begin{array}{c}n<m\\m_i=n_i\end{array}}{\sigma^3_m},\qquad
	\tilde{\sigma}^{1/2}_{n+h_i}=\sigma^{1/2}_{n+h_i}
		\prod_{\begin{array}{c}m<n+h_i\\m_i=n_i+1\end{array}}{\sigma^3_m},
\end{equation}
and we obtain the properties (\ref{eq:approximationalproperties}). The key is that for each node $m$ with $n<m,m_i=n_i$ the shifted point $m'=\tau_i m$ fulfils $m'<n+h_i$, thus, for each $\sigma^3_m$ in the first term we find a corresponding $\sigma^3_{m'}$ in the second term.

For our choice of $>$, the coefficients $\alpha^n_{n'}$ fulfill the following relations:
\begin{equation}
  \alpha^n_m = \alpha^{n+2h_i}_{m+2h_i};\ \ \hfill
	\alpha^n_{n+h_i} \alpha^{n+h_i}_{n+2h_i} = 1;\ \ \hfill
	\alpha^n_{n+h_i} \alpha^{n+h_i}_{n+h_i+h_j} = - \alpha^n_{n+h_j}\alpha^{n+h_j}_{n+h_i+h_j}.
	\label{eq:alpharelations}
\end{equation}

\subsection{Transformation of the lattice Dirac operator into staggered form}\label{staggering}

Now, the operator $H=\tilde{H}_D+mH_0$ appears to be a lattice Dirac operator.
Indeed, let's consider the evolution equation defined by $H$:
\begin{eqnarray}
	\label{eq:DiracInPsi}
	i\pd_t \psi_n   \ =\  [H,\psi_n  ]&=& \phantom{-}
		\sum_i \alpha^n_{n+h_i} (\psi_{n+h_i}^*-\psi_{n-h_i}^*)-m\psi_{n},\\
	\label{eq:DiracInPsiAdjoint}
	i\pd_t \psi_n^* \ =\  [H,\psi_n^*]&=& -
		\sum_i \alpha^n_{n+h_i} (\psi_{n+h_i}  -\psi_{n-h_i}  )+m\psi^*_{n}.
\end{eqnarray}
As a consequence of the relations (\ref{eq:alpharelations}), the evolution equations (\ref{eq:DiracInPsi}),(\ref{eq:DiracInPsiAdjoint}) give
\begin{equation}
	\pd^2_t\psi_n = \sum_i (\psi_{n+2h_i}-2\psi_n+\psi_{n-2h_i})-m^2\psi_{n}
	= -((\Delta_{2h}+m^2) \psi)_n,
	\label{eq:KleinGordonLattice}
\end{equation}
where $\Delta_{2h}$ is the lattice Laplace operator with doubled distance $2h_i$ -- a Laplace operator on a coarse lattice. 

The lattice Laplace operator $\Delta_{2h}$ acts independently on $2^d$ different sublattices. Let's distinguish these sublattices by introduction of $2^d$ different lattice functions enumerated by elements of $\k=(\k_1,\ldots,\k_d)\in\left\{0,1\right\}^d$. Using the denotation $*\psi_n=\psi^*_n$, we define
\begin{equation}
	\psi_\k(n) = *^{\k_1+\ldots+\k_d}\psi_n\qquad\textrm{on}\qquad n \equiv \k \;\;\textrm{mod}\;\; 2.
	\label{eq:def:psikappa}
\end{equation}

Each of the $2^d$ lattice functions $\psi_\k(n)$ is defined on a ``coarse lattice'' containing the nodes  of type $n_i = 2\tilde{n}_i+\k_i$ and lattice spacing $2h_i$. Now, the lattice operator $\Delta_{2h}$ acts as the simple Laplace operator on each of the $2^d$ functions $\psi_\k(n)$. In the continuous limit, each $\psi_\k(n)$ gives a function $\psi_\k(x)$ which fulfills the Klein-Gordon equation
\begin{equation}
	\pd^2_t\psi_\k(x,t) =(\sum_i \pd^2_i - m^2) \psi_\k(x,t) = 0.
	\label{eq:KleinGordonContinuous}
\end{equation}

The lattice Dirac equations (\ref{eq:DiracInPsi}),(\ref{eq:DiracInPsiAdjoint}) now establish a connection between these $2^d$ lattice fields. We can define now $2^d\times 2^d$ matrices $(\a^i)_\k^{\k'},\b_\k^{\k'}$ so that the original lattice equations (\ref{eq:DiracInPsi}),(\ref{eq:DiracInPsiAdjoint}) transform into
\begin{equation}
\begin{split}
	i\pd_t \psi_\k(n)   \ &=\  [H,\psi_\k(n)  ]\\&=\
		\sum_i -i(\alpha^i)_\k^{\k'} (\psi_{\k'}({n+h_i})-\psi_{\k'}({n-h_i})) + m\b_\k^{\k'}\psi_{\k'}(n)
		\label{eq:DiracLatticeDerived}
\end{split}
\end{equation}
on $n = \k \;\;\textrm{mod}\;\; 2$. Note that, because of the factor $*^{\k_1+\ldots+\k_d}$ in (\ref{eq:def:psikappa}), equation (\ref{eq:DiracLatticeDerived}) connects only the fields $\psi_\k(n)$, and it's adjoint only the $\psi^*_\k(n)$.

This equation is our lattice Dirac equation (\ref{eq:DiracLattice}) on the staggered lattice (\ref{eq:DiracStaggered}), but already in its quantized form, with anticommuting fermion operators $\psi_\k(n)$, and with a mass term.

\subsection{From spin fields to scalar fields}\label{spin2scalar}

Spin fields are already a much more classical object in comparison with the original fermion fields. But for our approach we need even more classical objects, namely real-valued fields.

Fortunately, this is not problematic at all. We can embed the spin field as an effective description of the lowest energy states of a scalar field with a \B-symmetric potential which gives two \B-symmetric vacuum states. For example, we can consider $\varphi^4$ theory in $\R^d$ with negative mass parameter $\mu^2$:
\begin{equation}
	\mathscr{L} = \frac{1}{2}((\pd_t\varphi)^2 - (\pd_i\varphi)^2)-V(\varphi) \;\textrm{with}\;
	V(\varphi)=-\frac{\mu^2}{2}\varphi^2+\frac{\lambda}{4!}\varphi^4.
	\label{eq:L:phi4}
\end{equation}
The two minima of the potential are $\varphi(x) = \pm \varphi_0$ with $\varphi_0 = \sqrt{\frac{6\mu^2}{\lambda}}$.

If the system is near $\varphi_0$, it is convenient to use the $\sigma$-variable $\sigma(x)=\varphi(x)-\varphi_0$ so that
\begin{equation}
	V(\sigma) = \frac{1}{2}(2\mu^2)\sigma^2+\sqrt{\frac{\lambda}{6}}\mu\sigma^3+\frac{\lambda}{4!}\sigma^4
	\label{eq:V:sigma}.
\end{equation}
This describes a scalar field with mass $\sqrt{2}\mu$ and some interactions.

Instead, we are interested only in the lowest energy states of this theory. Let's consider at first the simple case of dimension $d=0$, where QFT reduces to ordinary quantum theory. If we have energies much below $\mu$, only the two vacuum states $\Psi_\pm(\varphi)$ with $\langle \Psi_\pm|\varphi|\Psi_\pm\rangle \approx \pm\varphi_0$ are important. But the true eigenstates of energy are
\begin{equation}
	\Psi_{0/1}(\varphi)=\frac{1}{\sqrt{2}}(\Psi_+(\varphi) \pm \Psi_-(\varphi)).
	\label{eq:Psi}
\end{equation}
Between them, we have an energy gap of order
\begin{equation}
	\Delta = E_1-E_0 \sim \exp(-\int_0^{\varphi_0}{\sqrt{V(\varphi)-E_0}d\varphi}) \sim
	\exp(-\frac{\mu^3}{\lambda}).
	\label{eq:gap}
\end{equation}
With increasing $\mu$ the mass of the $\sigma$ field increases, but the energy gap $\Delta$ decreases exponentially. Without any conspiracy, this leads to two different domains: a high energy domain, with energies of order $\mu$, and a low energy domain, with energies of order $\Delta$, where the whole theory reduces to the two-dimensional space spanned by $\Psi_{0/1}$. Reduction to this subspace gives
\begin{subequations}
\begin{align}
\varphi &\quad\to\quad \varphi_0\sigma^1&\textrm{ with }\;\varphi_0&=\int\overline{\Psi}_0\cdot\varphi\Psi_1 d\varphi\\
\pi=\frac{\delta L}{\delta\dot{\varphi}} &\quad\to\quad \pi_0\sigma^2& \textrm{ with }\; \pi_0&=\int\overline{\Psi}_0\cdot\pd_{\varphi}\Psi_1 d\varphi\\
H&\quad\to\quad H_0-\frac{1}{2}\Delta \sigma^3&\textrm{ with }\; H_0&=\frac{E_0+E_1}{2}.
	\label{eq:reductionToPauli}
\end{align}
\end{subequations}
For dimension $d>0$, at least as long as the momentum $k$ is sufficiently small, we have a similar situation for each of the modes $\varphi(x)=\exp(ikx)\varphi$. For sufficiently large $\mu$, and suffiently low energies under consideration, the theory reduces to an effective theory where we have only two degrees of freedom for each mode. Effectively, the configuration space reduces from $\mathcal{F}(\Z^d,\R)$ to $\mathcal{F}(\Z^d,\B)$.

Last but not least, let's consider typical lattice theory interaction terms which may appear in the reduction for a Lagrangian of type (\ref{eq:L:phi4}). We consider lattice approximations where only neighbour nodes have nontrivial interaction terms. Let $n,n'=n+h_i$ be these neighbour nodes, $d=1$. One possibility is to use $\frac{1}{2}(\pi_n+\pi_{n'})$ to approximate $\pi(x)$ and $\frac{1}{h} (\varphi_n-\varphi_{n'})$ to approximate $\pd_i\varphi(x)$. Then, the reduction gives an effective Hamiltonian
\begin{equation}
	H = \frac{1}{2}(\pi^2 + c^2 (\pd_x\varphi)^2 +V(\varphi)) \to c_0+
c_1 \sigma^1_n\sigma^1_{n'} +
c_2 \sigma^2_n\sigma^2_{n'} +
c_3 \sigma^3_n
	\label{eq:HRtoHZ2}
\end{equation}
for some constants $c_i$. The lattice Dirac operator corresponds to $c_1 = -c_2 = 1$, thus, can be obtained in this scheme.

As a consequence of this quantization method for fermions, we obtain some analogon of a ``supersymmetric partner'' of the fermions. This partner can be very heavy without any conspiracy. At the current state of research, no indications about their masses can be given.

\subsection{Conclusion}

Given that our proposal to obtain fermionic fields via canonical quantization of real fields is a very drastic departure from conventional wisdom about fermion quantization, let's summarize here the results of this section:

\begin{itemize}

\item The connection between real fields and spin fields is conventional and unproblematic. The only unconventional element we have used here is to use a discretization in space but not time. 

\item  In one-dimensional space, for our choice of the order, the Hamilton operator does not require any approximation. Thus, for spatial dimension $1$, the problem of obtaining fermions from canonically quantized \B-field operators $\sigma^i_n$ on the lattice is solved exactly and completely. Once the question, if fermions may be obtained via canonical quantization of real fields, at least at a first look seems to be dimension-independent, this exact result for dimension $1$ seems to be a powerful argument for a positive answer.

\item The isomorphism between the operator algebra of spin field operators $\sigma^i_n$ and algebra of fermion operators $\psi_n,\psi_n^*$ is, in every dimension, and for every choice of the order, an exact isomorphism. It is only the Hamilton operator which has to be approximated in higher dimensions, and which depends on the choice of the order. But the definition of the quantum operator algebra is, certainly, the more important, fundamental part of quantization. Thus, the most important part of the quantization scheme is exact too.

\end{itemize}

A better justification of the choice of the order, as well as a evaluation of the error made during the approximation, have to be left to future research. Nonetheless, in the light of the exact results we have obtained, the general problem, if fermion fields may be obtained via canonical quantization of real fields, may be considered as solved.

\section{Some notes about gauge field quantization} \label{sec:quantumGaugeFields}

After the fermion quantization, the approach to quantum gauge fields is a second drastic departure from conventional wisdom: With the correction coefficients for lattice deformations, we propose to use, essentially, a non-gauge-invariant regularization for the chiral gauge fields. Moreover, we reject, in general (that means, even for the Wilson lattice gauge fields) the concept of BRST quantization in its current form.

In our condensed matter approach, the gauge fields have to be obtained as effective fields from more fundamental fields. In the case of the Wilson gauge fields, these will be degrees of freedom which describe the material between the cells, which effectively influences the connection between two neighbour cells, in such a way that it has to be compensated by the gauge transformation corresponding to the edge. Instead, the correction coefficients for lattice deformations are, essentially, degrees of freedom of the configuration of the cells. Thus, they have been already quantized by the quantization of the fermionic part. In above cases, the gauge fields do not have to be quantized themself, in a separate quantization procedure, but to be derived, as effective quantum fields, similar to phonons, from the more fundamental quantum theory.\footnote{
Let's note in this connection the approach of Volovik \cite{Volovik}, who uses examples of usual condensed matter, for example $^3He-A$, as analogies to fundamental physics.  Especially, he considers the appearance of effective gauge fields and gravitational fields in usual materials. His observations appear close to our approach. For example \cite{Volovik}: ``The $^3He-A$ analogy also suggests that all the degrees of freedom, bosonic and fermionic, can come from the initial (bare) fermionic degrees of freedom.''
}
How to do this has to be left to future research. Despite this, let's give here answers to some objections against our approach, which are based on the standard BRST approach to gauge field quantization.

A first objection against our construction of weak gauge fields in section \ref{weak} is that it presents a lattice regularization for chiral gauge field theory. But to obtain such a regularization is a famous problem of chiral lattice gauge theory \cite{GuptaLattice}, and there are various no-go theorems for such regularizations. But the regularization problem of chiral gauge theory is the problem to find a \emph{gauge-invariant} regularization. Our regularization has no exact gauge invariance on the lattice. Instead, we have only approximate gauge invariance -- the generators of the gauge group are associated with nontrivial lattice shifts. Thus, our regularization is not in contradiction with the various no-go theorems for regularizations with exact gauge invariance.

This answer leads, in a natural way, to a second objection. Last but not least, people have tried to find a regularization with exact gauge invariance not just for fun, but for a good reason -- to quantize chiral gauge fields. The problem is that the standard procedure to quantize gauge fields -- BRST quantization -- depends essentially on exact gauge invariance of the theory. Without exact gauge invariance, it fails miserably: What remains is a non-unitary theory. But this failure is a special problem of the manifestly Lorentz-covariant Gupta-Bleuer approach to gauge field quantization, which starts with an indefinite Hilbert space structure. Following a proposal of Gupta \cite{Gupta} and Bleuer \cite{Bleuer} for QED, manifest relativistic invariance is reached in the BRST approach using an unphysical ``big space'' with indefinite Hilbert metric. A physical interpretation of this big Hilbert space would lead to negative probabilities, which is nonsensical.  To get rid of the states with negative probability, restriction to an invariant subspace and factorization is used. But these operations depend on exact gauge invariance. If gauge invariance fails, the result is fatal for the whole approach.

But there is even a classical alternative -- the earlier approach of Fermi \cite{Fermi} and Dirac \cite{Dirac}, where the Hilbert space is definite, but Lorentz covariance is not explicit. Whatever may go wrong, the Hilbert space remains definite, and at least a probability interpretation of the results is possible. Another possibility is our approach, where quantization is done in terms of a more fundamental theory, which is based, of course, on a definite Hilbert space. Especially, in our model we have quantized the degrees of freedom of the cells (whose positions define the lattice, and, therefore, also the correction coefficients for lattice deformations) with classical canonical quantization. Then, the gauge fields are derived as effective fields, similar to phonons. In this approach, there is no place for an indefinite Hilbert space structure, and, therefore, a non-unitary theory, to appear. Let's note that, once we don't require exact gauge symmetry, the standard rejection of anomalous gauge fields is no longer justified. 

The rejection of the manifestly Lorentz-covariant approach leads to another objection: It has not been introduced without reason too. Manifest Lorentz covariance, on one hand, simplifies computations. This is, obviously, not a decisive argument. More serious is that, in our approach, we do not have relativistic invariance. Indeed, our construction from the start violates Lorentz covariance, and in many different ways: First, we handle time and space in different ways, having a lattice only in space. Then, even a spacetime lattice would violate the symmetry of the continuous limit. Moreover, the operators $\sigma_{ij}$, for spatial spinor rotations on our staggered lattice, are nonlocal, and, therefore, do not define an exact representation of the algebra $\mathfrak{su}(2)$.\footnote{This violation of Lorentz invariance appears also for four-dimensional staggered fermions.} An approach, which violates Lorentz invariance on the fundamental level, has to explain how it will be recovered in the large distance limit.

Fortunately, this question has been, at least partially, addressed by the derivation of the Einstein equivalence principle (which includes local Lorentz covariance) in our theory of gravity \cite{GLET}.  On the other hand, the connection between the postulates of this theory of gravity and the approach described here remains unclear. Especially, our fermion quantization procedure does not lead to a unique ``speed of light'' for different fermions. For the solution of this question, a result of Chadha and Nielsen \cite{Chadha} may become important: They considered massless electrodynamics with different metrics for the left-handed and right-handed fermions; their model thus violates Lorentz invariance. They found that the two metrics converge to a single one as the energy is lowered. Thus, in the low-energy corner Lorentz invariance becomes better and better.

\section{What about the masses?}

Up to know, only the massless case has been considered. What can be said about the masses? 

The approach itself does not promise a great ability to compute the masses from first principles.  Indeed, if there is a fundamental ether, this ether will have its own material parameters, and the masses of the effective particles will plausibly depend on these unknown material parameters of the ether. So there will be a lot of unknown parameters which can be used to fit the observed particle masses. 

But this does not mean that the approach remains completely silent about the mass terms. The masses of the standard model have some quite specific qualitative characteristics which require an explanation. 

In particular extremely large mass differences require explanations. Especially there are the following points:

\begin{itemize}

\item The masslessness of the gluons and the photon, in distinction with the non-zero masses of the weak gauge bosons. 

\item The extremely small masses of the neutrinos in comparison with the masses of the other fermions.

\item The smallness of the CP-violating terms in the mass matrix.

\end{itemize}

So is there some hope to explain these properties? The following considerations provide at least some ideas.

\subsection{Neutrinos as pseudo-acoustic phonons}

Remember the association between the right-handed neutrinos and translational symmetry: It is a consequence of the postulate of Euclidean symmetry. While rotational symmetry leads to identical gauge actions on all three generations, the consequence of translational symmetry was the existence of an inert direction in each generation. Once the only inert particles are right-handed neutrinos, the translational direction has to be associated with the right-handed neutrino sector. 

But this is not the only consequence of translational symmetry. Another one is standard knowledge in condensed matter theory, and leads to the notion of ``acoustic phonons'' being different from ``optical phonons'': Acoustic phonons have no mass gap, while optical phonons have a mass gap (see, for example, sec. 3.7 of \cite{Taylor}). This distinction is explained by translational symmetry: Essentially, the acoustic phonons are the Goldstone bosons related with Galilean symmetry, which is broken by the velocity of the background \cite{Schakel}. As a consequence, we have exactly three acoustic phonons -- the Goldstone bosons of the three-dimensional group of Galilean boosts. All other phonon modes have a mass gap -- one needs some minimal energy to excite them.

Now, the acoustic phonons themself are density waves, and density $\rho$, velocity $v^i$ and the stress tensor $\sigma^{ij}$ have been identified in the model with the gravitational field. Thus, the acoustic phonons themself are gravitons\footnote{More accurate, they are among the additional degrees of freedom of the gravitational field associated with the change of the preferred coordinates in the non-covariant theory of gravity proposed in \cite{GLET}, thus, decouple in the GR limit of this theory. Thus, they are not the quadrupol waves of the GR limit.} And, in nice correspondence with the prediction that the acoustic phonons have to be massless, the mass of the graviton is yet below the level of detection and much smaller than all masses of fermions.

Nonetheless, the neutrinos are associated with translations of a subsystem of the medium -- the lattice of cells. This subsystem interacts with the other parts of the medium -- the material between the cells. So we have no exact translational symmetry for the neutrinos. But, if we assume that the interaction with the surrounding medium is sufficiently weak, there may be an approximate translational symmetry for the subsystem of the cells. And such an approximate symmetry leads to corresponding Goldstone particles with low masses.

But this is exactly what we need: The masses of the three neutrinos are many orders smaller than those of the other SM fermions, but not zero. This mass difference is large enough to require an explanation, and the standard way to explain such small masses is an approximate symmetry -- exactly what we have found, with the translational symmetry of the lattice of cells.

With similar arguments one may hope to explain the higher masses of the quarks in comparison with the leptons, and the higher masses of the upper in comparison with the lower quarks: the higher masses correspond to modes which are in some sense more far away from the translational symmetry. But this is, of course, much too vague, and clearly requires future research. 

\subsection{Why are the gluons massless but weak bosons massive?}

The association between masslessness and symmetry may be used, as well, to explain the masslessness of the gluons, while the weak gauge bosons are massive. 

The symmetry which can be used here is the obvious candidate -- the gauge symmetry. Remember that the strong gauge fields have a different representation in the lattice model -- they are implemented by an analogon of a Wilson gauge field.  And, as a consequence, there exists an \emph{exact} lattice gauge symmetry for the strong interactions $U(3)_c\supset SU(3)_c$ which contains all the gluons. 

The situation is different for the weak gauge bosons.  In the lattice model, they appear as effective fields which describe geometric deformations of the lattice. While there is a pointwise lattice gauge transformation in the case of the Wilson-like gauge fields, no such exact gauge symmetry exists for these effective fields. 

And, so, the standard correspondence between symmetries and Goldstone bosons leads to the quite natural prediction that gluons have to be massless, while weak bosons will be massive. 

\subsection{Why is the phonon massless?}

So the situation seems quite easy for gluons and weak bosons -- we have an ideal correspondence between exact lattice gauge symmetry and masslessness of the corresponding gauge bosons. 

The case of the photon is more complicate, because it is a combination of the strong $U(1)_B$ part, which has an exact lattice gauge symmetry, and a weak part, which does not have such a symmetry.  

We know the result: The photon is massless too. So it seems reasonable to hope for some mechanism which transforms the exact lattice gauge symmetry of the $U(1)_B$ field into the masslessness of the photon. 

And it seems that this hope is not completely unreasonable. The very point of the Goldstone theorem is that if there is a symmetry, then there will be a corresponding massless particle.  If it cannot be the $U(1)_B$ field, because it violates the postulate of ground state neutrality, it has to be a different field, and the best candidate would be the field closest to $U(1)_B$ which is part of the SM.  

But this is, in a quite obvious way, the electromagnetric field.  

Indeed, it is the only candidate among the remaining SM fields which is a vector field, thus, preserves an essential qualitative property of $U(1)_B$. 

And there is also another similarity, which becomes important if we look at the details of the gauge action on the level of the gauge groups itself. In the formulas above, I have written, in a sloppy way, expressions like $U(3)_c \cong SU(3)_c \times U(1)_B$.  But this is correct only on the level of the Lie algebras. On the group level, the correct formula is a slightly different one:
\begin{equation}
U(3)_c \cong (SU(3)_c \times U(1)_B)/\Z_3.
\end{equation}
The difference is usually not very important, because a representation of $U(3)$ also defines a representation of $SU(3)\times U(1)$. But not every representation of $SU(3)\times U(1)$ defines also a representation of $U(3)$. For this, it is necessary that the element $e^{2\pi i/3}\in U(1)$ is represented in the same way as the diagonal matrix $e^{2\pi i/3}E \in SU(3)$.  

Now, the exact lattice gauge symmetry, which we want to use to explain the masslessness of the photon, is an $U(3)$ symmetry. Even if somehow deformed, once can expect that it has to remain an $U(3)$ symmetry. We want to apply this to the representation of $SU(3)_c\times U(1)_\g$. But is this representation of $SU(3)\times U(1)$ also a representation of $U(3)\cong(SU(3)\times U(1))/\Z_3$? 

It is. And I think this point is some additional hint that the idea to consider the action of $SU(3)_c\times U(1)_\g$ as a deformation of the $U(3)_c$ action, which allows to explain the masslessness of the photon too, is a reasonable one. 

So there seems to be a natural route for an explanation of the masslessness of the $SU(3)_c\times U(1)_{em}$ part of the SM gauge group. This part should be interpreted as a deformation of the $U(3)_c$ group action which has an exact lattice gauge symmetry. It fulfills the following properties:
\begin{enumerate}
\item The symmetry group of the exact gauge symmetries itself remains unchanged -- it is $U(3)\cong (SU(3)\times U(1))/\Z_3$.
\item The action of the group remains to be a vector action.
\item The action is among the ones which are compatible with all other requirements, in particular with charge neutrality of the ground state.
\end{enumerate}
And it is easy to see that these properties already uniquely identify the $U(1)_{em}$ action as the result of the deformation. Indeed, all other $U(1)$ subgroups of the maximal possible group $G_{max} \cong S(U(3)_c\times U(2)_L\times U(1)_R)$ are chiral (violating condition 2) or do not commute with $SU(3)_c$ (violating condition 1).

So, even if the explanation of the masslessness of the photon requires some additional research, and in particular a mechanism of the deformation of the action of the symmetry, there is at least a reasonable hope to obtain such an explanation. 

\section{Discussion}

Many interesting questions have to be left to future research. This includes:

\begin{itemize}

\item Symmetry breaking;

\item The search for a Hamilton operator for a general configuration of cells, which would allow the derivation for other regular crystallographic lattices as well as for lattices with deformations and defects;

\item The large distance limit, especially renormalization group equations; 

\item The connection between our lattice model for the SM and the theory of gravity presented in  \cite{GLET}. They are conceptually and metaphysically compatible, but mathematically yet unrelated; In the large distance limit of our lattice, we will obviously have notions like density $\rho$, average velocity $v^i$ and  some stress tensor $\sigma^{ij}$, which follow continuity and Euler equations. Nonetheless, the open question is how to obtain the corresponding special Lagrange formalism postulated in this theory of gravity.

\end{itemize}

Especially symmetry breaking promises to be interesting. First, we need it. The \E\/ action does not define a symmetry of the SM: The fermionic mass terms clearly violate the rotational symmetry between the three generations. Moreover, the whole construction of section \ref{fermionQuantization}, which creates effective \B-fields from the original \R-fields, violates translational symmetry: We cannot add real constants to \B-valued fields. Thus, to obtain the Hamiltonian of the SM, even to obtain fermion fields at all, we have to break \E\/ symmetry. 

On the other hand, some points suggest large differences between the SM symmetry breaking scheme and the symmetry breaking we need in our approach:

\begin{itemize}
\item The idea that the gluons and the photon is massless because they are connected with the exact lattice gauge symmetry originally connected with the $U(3)_c$, while the weak bosons are massive because there is no such exact lattice gauge symmetry, invalidates the original necessity for symmetry breaking -- the aim to give the weak bosons a mass.

\item A standard argument for a separate Higgs sector is that spatial isotropy is unbroken. But in our model, spatial isotropy is broken by the mass terms. Thus, this argument in favour of a separate Higgs sector fails. The question appears if we need a separate Higgs sector at all. The application of Ockham's razor -- don't introduce Higgs fields without necessity -- suggests that there will be no Higgs sector at all. 
\end{itemize}

Thus, we need symmetry breaking, but for very different reasons, and of a different symmetry. Therefore we expect the symmetry breaking in our model to be very different from the SM scheme. The only common thing will be, possibly, the general idea of symmetry breaking, and that it has to give fermions mass.

Despite these open questions, our lattice model, being of surprising simplicity, already has a place for all standard model particles observed until now. The fermionic content of the SM is predicted exactly, with all three generations. Already this part, taken alone, gives our model large explanatory power: Even for small natural modifications (two or four instead of three generations, two or four instead of three colors, electroweak singlets, triplets or quartets instead of doublets) where would be no corresponding condensed matter model of similar simplicity.

Then, the maximal gauge group compatible with few simple postulates gives the SM gauge group exactly. And even if one weakens the requirement by allowing anomalous gauge fields, there is room for only one additional gauge field $U(1)_{U}$. This is a quite impressive result, given that the maximal possible unitary gauge group on the fourty eight Weyl fermions of the SM is the $48^2$-dimensional group $U(48)$, which is reduced to the $12$-dimensional SM gauge group $SU(3)_c\times SU(2)_L\times U(1)_Y$ itself resp. its $13$-dimensional extension $S(U(3)_c\times U(2)_L\times U(1)_R)$.

The comparison of these results with those of other approaches, like string theory, will be left to the reader.

Last but not least, the model is conceptually compatible with a metric theory of gravity with GR limit presented in \cite{GLET}. Thus, it gives also hope for a development of a ``theory of everything'', which unifies the SM fermions and gauge fields with gravity.

\begin{appendix}

\section{The Dirac operator on the exterior bundle}\label{DiracCurved}

We give here the basic formulas for the Dirac operator in the exterior bundle (see, for example, \cite{Pete}). The exterior bundle or de Rham complex $\Lambda = \sum_{k=0}^d \Lambda^k$ consists of skew-symmetric tensor fields of type $(0,k), 0\le k \le d$, which are usually written as differential forms
\begin{equation}
\psi = \psi_{i_1\ldots i_k} dx^{i_1}\wedge \cdots \wedge dx^{i_k} \in \Lambda^k
\end{equation}
The exterior bundle $\Lambda$ has dimension $2^d$ in the d-dimensional space.  The most important operation on $\Lambda$ is the external derivative $d:\Lambda^k\to\Lambda^{k+1}$ defined by
\begin{equation}
(d\psi)_{i_1\ldots i_{k+1}}=\sum_{q=1}^{k+1}\frac{\partial}{\partial x^{i_q}}
 (-1)^q \psi_{i_1\ldots \hat{i}_q\ldots i_{k+1}} 
\end{equation}
where $\hat{i}_q$ denotes that the index $i_q$ has been omitted. It's main property is $d^2=0$.  In the presence of a metric, we have also the important $\star$-operator $\Lambda^k\to\Lambda^{d-k}$:
\begin{equation}
(\star\psi)_{i_{k+1}\ldots i_d} = \frac{1}{k!} \varepsilon_{i_1\ldots i_d} 
g^{i_1j_1} \cdots g^{i_kj_k}\sqrt{|g|}\psi_{j_1\ldots j_k}
\end{equation}
with $\star^2 = (-1)^{k(d-k)}\mbox{sgn}(g)$.  This allows to define a global inner product by
\begin{equation}
(\phi,\psi) = \int \phi \wedge (\star\psi) = \int \psi \wedge (\star\phi)
\end{equation}
It turns out that the adjoint operator of $d^*: \Lambda^k\to\Lambda^{k-1}$ of $d$ is
\begin{equation}
d^* = (-1)^{kd+d+1} \star d \star
\end{equation}

Note that the expressions for $\star^2$ and $d^*$ depend on the order of the form $k$, which is not nice. But a minor redefinition of the $\star$ operator allows to solve this problem. For the operator
\begin{equation}\label{astdef}
 \ast = i^{k(d-k)}\star
\end{equation} 
the resulting expressions no longer depend on $k$:
\begin{equation}
 \ast^2 = \mbox{sgn}(g), \qquad d^* = (-i)^{d+1} \ast d \ast 
\end{equation} 

In this general context we can define the Laplace operator as
\begin{equation}
\Delta = d d^* + d^* d.
\end{equation}
Then, the Dirac operator (as it's square root) can be defined as
\begin{equation}
D = d+d^*,
\end{equation}
so that $\Delta=D^2$. Indeed, we have $d^2=0$ as well as $(d^*)^2=0$.

\subsection{Discretization of the Dirac operator}\label{DiracLatticeCurved}

The special geometric nature of the exterior bundle allows to define a nice doubler-free discretization of the Dirac equation on a general cell complex. Such a cell complex consists of cells $c_i$ of dimension $k=dim(c_i)$, which are embeddings of the $k$-dimensional unit cube $I^k$ into the manifold so that the image of the boundary is part of the image of lower-dimensional cells of the complex, and the image of all cells of the cell complex covers the whole manifold. 

On such a cell complex, $k$-dimensional differential forms are represented on the lattice by their integrals over the $k$-dimensional cells $c_i$ of the cell complex:

\begin{equation}
 \Psi \to \{\psi_i\}, \psi_i = \int_{c_i} \Psi
\end{equation} 

The external derivative defines in a similar natural way a derivative for functions on the mesh, with the same most important exact property $d^2=0$.

For the definition of the $\star$-operator we need a metric and a dual mesh.  A metric $g_{\mu\nu}$ on the manifold defines in a natural way for every cell $c_i$ it's area $a_i=a(c_i)> 0$.  For a triangulation on Euclidean background, the values $\a_i$ depend on each other. But in the general case they may be considered as independent variables, which approximate the metric on the cell complex. In the following we consider them as given and defining the metric.

A dual mesh is a mesh with cells $\hat{c}_i$ so that for each cell $c_i$ of the original mesh with dimension $k$ we have a corresponding ``dual cell'' $\hat{c}_i$ of dimension $d-k$ which intersects only the cell $c_i$, in a single point, orthogonally and with positive intersection index. The metric defines the areas $\hat{a}_i$ of the dual cells in a similar way. Now, the lattice Hodge $\star$-operator may be defined as

\begin{equation}
\star \psi_i = \frac{\hat{a}_i}{a_i} \hat{\psi}_i
\end{equation}

and maps functions on the mesh to functions on the dual mesh.  Note that the dual of the dual mesh has the same cells as the original mesh, but possibly with different orientations. Therefore, for $\star^2$ we obtain an additional factor  $(-1)^{k(d-k)}$ as in the continuous case.

Thus, we can define the exterior derivative as well as the Hodge $\star$ operator on the lattice preserving their algebraic properties $d^2=0$, $\star^2=(-1)^{k(d-k)}$. As a consequence, the remaining part of the theory can be transferred on the lattice too.

It is a general consequence of the geometric character of the continuous Dirac equation as well as its lattice discretization that the lattice discretization does not have doublers. See, for example, \cite{Banks,Becher,BecherJoos,Rabin}.

\end{appendix}

\end{document}